\documentclass[twocolumn,showpacs]{revtex4}

\usepackage{graphicx}

\usepackage{amsfonts}

\usepackage{amssymb}

\usepackage{array}
\usepackage{graphicx}
\usepackage{pstricks}
\usepackage{color}
\usepackage{xcolor}

\usepackage{subfigure}
\usepackage{amsmath}

\usepackage{latexsym}

\begin{document}
%

\title{The weak bound state with the non-zero charge density as the LHC 126.5 GeV
state\footnote{Published in: J. Phys. Part. Nuclei (2016) 47: 838-862.}}\author{Jacek Syska  }\email[E-mail: ]{jacek.syska@us.edu.pl}
\affiliation{Department of Field Theory and Particle Physics,
\\Institute of Physics, University of Silesia,  Uniwersytecka 4,
40-007 Katowice, Poland}



\begin{abstract}
\vspace{-2mm}
The self-consistent model of classical field
interactions formulated as the counterpart of the quantum
electroweak model leads to
homogeneous boson ground state solutions in presence of non-zero extended fermionic charge density fluctuations.
Two different types of electroweak configurations of fields are analyzed. The first one has non-zero electric and weak charge fluctuations. The second one is electrically uncharged but weakly charged. Both types of configurations have two physically interesting solutions which possess masses equal to 126.67
GeV at the value of the scalar fluctuation potential parameter $\lambda$ equal to $\sim 0.0652$.
The spin zero electrically uncharged droplet
formed as a result of the decay of the charged one is interpreted as the $\sim 126.5$ GeV state found in the Large Hadron Collider (LHC) experiment. (The other two
configurations
correspond to solutions with masses equal to 123.7 GeV and
$\lambda$ equal to $\sim 0.0498$
and thus the algebraic mean of the masses of two central solutions, i.e.,  126.67 GeV and 123.7 GeV, is equal to 125.185 GeV.)
The problem of a mass of this kind of droplets will be considered on
the basis of the phenomenon of the screening of the fluctuation of
charges. Their masses are found in the thin wall approximation.
%
%
\end{abstract}

\vspace{-1mm}


\maketitle



{\footnotesize
\tableofcontents
}

\normalsize

\section{Introduction}

\label{Introduction}

\vspace{-3mm}

In \cite{Dziekuje_Jacek_nova_2}, the non-linear self-consistent model of classical field interactions in the ``classical counterpart of the
electroweak Glashow-Salam-Weinberg'' (CGSW) model was proposed. Homogeneous boson ground
state solutions in this model in the presence of non-zero extended
fermionic charge density fluctuations were reviewed and
fully reinterpreted in order to make the theory with non-zero charge
densities \cite{JacekManka} coherent as, unfortunately, the language in \cite{JacekManka} uses both quantum field theory (QFT) concepts and the classical charge distributions. The model concerns the bound
states of the matter of these fluctuations inside one droplet of
fields.
Because of the Pauli exclusion principle, only one or (for the sake of opposite projections
\cite{stochastic_projection_Grossing,stochastic_projection_Keppeler}, \cite{Dziekuje_za_EPR-Bohm,Dziekuje_za_skrypt} of the spin) two fermionic fluctuations in one droplet can occupy at their lowest energy state. Unless other quantum numbers are assigned to these fluctuations, the consecutive fermionic fluctuations can eventually occupy their higher energy states.
%
%
Concerning the phenomenon of the screening of the fluctuation of charges inside one droplet, we face the problem of the mass of this kind of droplet. The phenomenon of the gamma transparency of the electrically uncharged configuration of fields in the droplets in the reference to gamma bursts was previously pointed out in \cite{MarekJacek}. Below, the Schr\"{o}dinger-Barut background of the model is given.
\\
The analyzed CGSW model is not a modification of the quantum GSW model \cite{Arbuzov-Zaitsev}. For instance, the configurations of fields are not the structures of QFT; most particularly the ground
state is not the QFT vacuum state. Hence, the
argument against ``a non-zero vacuum expectation value'' is not
relevant here since in the body of the self-consistent field
theory, a structure like this does not exist at all.
Unlike QFT, the self-consistent field theory (SCFT)
deals with continuous charge densities and continuous charge
density fluctuations as the basic concept \cite{Dziekuje_Jacek_nova_2,Dziekuje_za_self-consistent}.
\\
In order to present the idea of the ground field in a broader
context, let us draw our attention to the Lagrangian
density ${\cal L}$ of electromagnetism, which serves as an example for  introducing the {\it ground field} notion in terms of the self-consistent theory only
\begin{eqnarray}
\label{lagr-el} {\cal L} = \bar{\Psi} ( \gamma^{\mu} i
\partial_{\mu} - m ) \Psi + J^{\mu} \, A_{\mu} - \frac{1}{4} \,
F_{\mu \nu} \, F^{\mu \nu} \, , \nonumber
\end{eqnarray}
where $J^{\mu} = - e \bar{\Psi} \gamma^{\mu} \Psi $ is the
electron current density fluctuation and $A_{\mu} $ is the total
electromagnetic field four-potential $A_{\mu} = A^{e}_{\mu} +
A^{s}_{\mu} $, where the superscript $e$ stands for the external
field and $s$ stands for the self-field adjusted by the radiative
reaction to suit the electron current and its
fluctuations (see \cite{Barut-1,Barut-2,Barut-3,Barut-4}, \cite{B-Nonlinear-1,B-Nonlinear-2}).
Then, in the minimum of the corresponding total
Hamiltonian, the solution of the equation of motion for
$A^{s}_{\mu}$ is called the {\it electromagnetic ground
field}.
\\
In this paper, the term {\it boson ground field} is used for the solution of equations of motion for a {\it boson field} in the ground state of the whole system of fields (fermion fluctuations, gauge bosons, scalar fluctuation) that are under consideration. This boson field is a self-field  (or can be treated as one) when it is coupled to a source-``basic'' field. In general, the term ``basic'' field means a wave
function that is proper for a fermion (fluctuation), a scalar (fluctuation) or a dilatonic field \cite{Dziekuje_CI_PANIE_JEZU_CHRYSTE,Dziekuje_za_neutron} and, although not in this paper,
a charged or heavy boson (which in this case plays simultaneously the role of both the basic and ground field). \\
The above mentioned concept of a wave function and the Schr\"{o}dinger wave equation is dominant in the nonrelativistic physics of atoms, molecules and condensed matter \cite{Sakurai}.
In the relativistic quantum theory, this notion has been largely
abandoned in favor of the second quantized perturbative Feynman
graph approach, although the Dirac wave equation is still used
for the approximation of some problems.
\\
What Barut and others did was to extend the Schr\"{o}dinger's
``charge density interpretation''  of a wave function (e.g. the electron is the classical distribution of charge) to a ``fully-fledged'' relativistic theory. They successfully implemented
this ``natural (fields theory) interpretation'' of a wave function
with coupled Dirac and Maxwell equations
(for characteristic boundary conditions) in many specific problems.
But the ``natural
interpretation'' of the wave function can be extended to the
Klein-Gordon equation \cite{Dziekuje_CI_PANIE_JEZU_CHRYSTE,Dziekuje_za_neutron} coupled to the Einstein field equations, thus being a rival for quantum gravity in its second quantization form. In the case of the QFT models, the second quantization approach
is connected with the probabilistic interpretation that is inherent in the quantum theory, whereas the classical field theories and the
``natural interpretation'' of the wave function together with the
self-field concept are in tune with the deterministic
interpretation forming a relativistic SCFT.
\\
Thus, depending on the model, the role of a self-field can
be played by e.g. the electromagnetic field
%
%
\cite{bib_B-K-1,relativ_Lamb,ion-collisions}, \cite{spontaneous-1,spontaneous-2}, \cite{bib_B-D,Casimir}, boson $W^{+}-W^{-}$ and $Z$
ground-field (as below in this paper)  \cite{Dziekuje_Jacek_nova_2,JacekManka} or by the gravitational field (metric tensor) $g_{\mu \nu}$ \cite{Dziekuje_CI_PANIE_JEZU_CHRYSTE,Dziekuje_za_neutron}. The ``basic'' field that is proper for a particular matter source is the dominant factor in the existence of self-fields. \\
When the values of masses of fundamental fermionic, scalar and bosonic fields have to be taken as the external parameters of the model, then in SCFT
%
%
the basic fields are in fact interpreted as fluctuations \cite{Jaynes-1,Jaynes-2,Jaynes-3}, \cite{Milonni} (of the total basic fields) and the self-fields are coupled to the fluctuations only. The conjecture is that if all fluctuations are identical to their total basic fields, then the solution is fully self-consistent and the masses of all fields should appear in the result of the solution of the coupled partial differential equations that characterize the system \cite{Frieden-1,Frieden-2,Frieden-3,Frieden-4,Frieden-5}, \cite{Dziekuje_za_skrypt,Dziekuje_za_models_building}.
In \cite{Frieden-1,Frieden-2,Frieden-3,Frieden-4,Frieden-5} it was shown that the structural information of the system  \cite{Dziekuje_informacja_2,Dziekuje_informacja_1,Dziekuje_za_EPR-Bohm} is, in the case of the scalar field, proportional to its squared rest mass.
The (observed) {\it structural information principle} put upon the system means that the analyticity requirement of the log-likelihood function  of the system \cite{Dziekuje_informacja_2,Dziekuje_za_EPR-Bohm} is used. The coupled set of self-consistently solved
partial differential equations
arises when
the {\it variational information  principle}, which minimizes the total physical information of the system \cite{Frieden-1,Frieden-2,Frieden-3,Frieden-4,Frieden-5}, \cite{Dziekuje_informacja_2} is also put upon the
system. (In the analyses,
the Rao-Fisher metricity of the statistical space \cite{Amari-Nagaoka_book}  of the system is used \cite{Dziekuje_za_channel,Dziekuje_za_EPR-Bohm}.) \\
If only some of the fluctuations are identified with their
total basic fields, then all masses of the fundamental fields remain among the parameters \cite{Dziekuje_za_channel} that (at least at some value of the energy) are to be estimated from the experiment.
\\
In accordance with the statement above, a model of bound states of
fluctuations (index~$f$) was constructed \cite{Dziekuje_Jacek_nova_2}.
The new, electrically and/or weakly {\it charged physical configuration} lies in the minimum of the effective
potential of the scalar field fluctuation $\varphi_{f}$ at the value $\varphi_{f} = \delta$, which is calculated self consistently from the Lagrangian of the CGSW model.
In the model, the scalar field $\varphi$ exists inside the droplet of the configuration of fields only. It is the only one (inside the droplet) to which its fluctuation
$\varphi_{f} \equiv \varphi$ is possibly  equivalent (possibly, as this paper neither proves nor disproves it). In fact, it could be an  effective one, e.g. the superposition of other fundamental fields or their fluctuations.  \\
%
%
Thus, from now on, the symbols $\varphi_{f}$, $L_{f}$, $R_{f}$, respectively, denote the fluctuation of the scalar field and a doublet of  left-handed or a singlet of right-handed fluctuations of fermionic fields, respectively, and not the global fields.
In agreement with the above explanations of the self-consistent approach,  fields in a doublet $L_{f} = {\nu_{f  L} \choose \ell_{f  L}}$ and a singlet $R_{f} = (\ell_{f  R})$ are wavefunctions, where $\ell_{f}$ and $\nu_{f}$ signify a leptonic fluctuation $\ell$ and a fluctuation of its neutrino $\nu$, respectively. Thus fields in $L_{f}$ and $R_{f}$ are not connected with the interpretation of the corresponding full (global) charge density distributions for particles in the doublet $L$ and singlet $R$, as it is for fields ruled by the original linear Dirac equation. Instead, they are associated with the distributions of the charge density {\it fluctuations of} fields in the doublet $L$ and singlet $R$ that are ruled by the coupled Dirac-Maxwell equations, similar to that found in Barut's case.
Therefore, $j_{f \, Y}^{\, \mu}$ and $j_{f}^{\, a \mu}$,  $a = 1,2,3$
are the continuous matter current electro-weak density fluctuations extended in space (and not operators of QFT with point-like charges).
In order to simplify the calculations, the mass $m_{f}$ of
any fermionic fluctuation
%
%
is neglected (see Eq.(\ref{Pdz_19})).
\\
In Section~\ref{basic solutions} the effective potential for the ``boson ground fields induced by matter sources'' configuration (hereafter, I will call it the bgfms configuration) and the general algebraic equations that follow from the field equations of motion for the fields on the
ground state inside the droplet are presented. They form the screening condition of the fluctuation of charges. Such quantities as the observed charge density fluctuations are also determined. In Section~\ref{electric} the numerical results for the electrically and weakly charged bgfms (EWbgfms) configuration are presented along with the calculations of the mass of its droplet in the thin wall approximation. Section~\ref{neutral} is devoted to an analysis of the weakly charged bgfms (Wbgfms) configuration and its stability for the sake of both the weak charge density fluctuation and $\lambda$ parameter (which is the parameter of the scalar fluctuation potential). In Section~\ref{intersection} the intersections of the $\lambda$-functions of the mass of the droplet for the electrically charged (i.e. EWbgfms) and electrically uncharged (i.e. Wbgfms) configurations are analyzed.
%
%
Two
of such pairs of bgfms configurations are found and analyzed: one with a mass equal to 123.7 GeV and the other with 126.67 GeV. Then, the Wbgfms configuration with a mass equal to 126.67 GeV is interpreted as the state found in the LHC experiment  \cite{cms2,atlas2} (the Wbgfms configuration with a mass equal to 123.7 GeV is also considered).
Also, in Section~\ref{intersection} the decay and gamma transparency of the Wbgfms configuration are described. After the Conclusions, in Appendix~1 the Table with some quantum numbers of fields in the $SU_{L}(2) \times U_{Y}(1)$ CGSW model are given. In Appendix~2 the field equations for the gauge self-fields and the scalar field fluctuation in CGSW model with continuous matter current density fluctuations are given. The calculations below are in the ``natural units'' $\hbar=c=1$.

\section{Boson ground state solutions}

\label{basic solutions}

In the CGSW model the
Lagrangian density for the fluctuations and self-fields coupled to them with the hidden $SU_L(2) \times U_Y(1)$ symmetry is as follows
\begin{eqnarray}
\label{lagrangian}
\!\!\!\!\!\!\!\!\!\!\!\!\!\!\! {\cal L}_{f} &=& - \frac{1}{4}
F^a_{\mu \nu}F^{a \mu \nu} - \frac{1}{4} B_{\mu\nu}B^{\mu\nu} +
(D_{\mu} \Phi_{f})^{+}D^{\mu}\Phi_{f} \nonumber \\
&-& \lambda (\Phi_{f}^{+} \Phi_{f} -
\frac{v^{2}}{2} )^{2} + {\cal L}_{f}^{\;f} \, ,
\end{eqnarray}
where ${\cal L}_{f}^{\;f}$ is the fermionic part of the fluctuation
sector
\begin{eqnarray}
\label{lagrangian fermionic}
\!\!\!\!\!\!\!\!\!\!\!\!\!\!\! {\cal L}_{f}^{f} &=& i \bar{L} \gamma^{\mu}
\nabla_{\mu} L_{f} + i \bar{R}_{f} \gamma^{\mu} \nabla_{\mu} R_{f} \nonumber \\
&-& \sqrt{2} \, \frac{m_{f}}{v}(\bar{L}_{f  L} \Phi_{f} \, R_{f} + \;h.c.) \;\; .
\end{eqnarray}
Here, $v=246.22 \; {\rm GeV}$ \cite{particle_data} and $\lambda \neq 0$ is the constant parameter of the scalar fluctuation potential, whose  value will be established later on. To simplify the calculations, we
neglect the mass $m_{\ell_{f}}$ of the fermionic fluctuation.  \\
%
%
The fields inside the bgfms droplet are either the classical fluctuations of fields or classical self-fields and in this paper they are treated as such. Because the formalism for the self-consistent treatment of the quantum fields operators is not known, therefore the fields of the self-consistent approach are not the ones of a quantum field theory origin. The same is true for the quantum fluctuation fields operators.
This concerns the scalar fluctuation doublet and all fermionic fluctuations and bosonic self-fields inside the bgfms configuration.
%
%
Moreover, both the bosonic self-fields and the scalar and fermionic fluctuations that compose the bgfms configuration are not directly observed. What is observed is the droplet of the bgfms configuration. In this respect, the clarifying (only) similarity is to think of the neutron as a kind of configuration of fields. It is hard to prove that it consists of a proton and an electron (although see
\cite{Santilli_1,Santilli_2}).
Similarly, it would be risky to call the fermionic fluctuation inside the droplet, e.g. a particular lepton fluctuation, although in the CGSW model the field fluctuations inside the droplet are granted the  $SU_L(2) \times U_Y(1)$ quantum numbers (see Table in Appendix~1). For example, the electrically charged EWbgfms
%
%
state found in Section~\ref{intersection} has the $SU_L(2) \times U_Y(1)$ quantum numbers of
the fermionic fluctuation(s),
which are the same as the numbers
of the positron. Also, the scalar fluctuation potential  $\lambda (\Phi_{f}^{+} \Phi_{f} - \frac{v^{2}}{2} )^{2}$ in the CGSW model is the one for the classical scalar field {\it fluctuation} $\Phi_{f}$ that exists inside the bgfms configuration only and not for the Higgs field. In conclusion, the CGSW model is the one of the fluctuations of basic (scalar or fermionic) fields and the self-fields coupled to them. The scalar or fermionic fluctuations can be the objects different than the ones known from, e.g. the scattering experiments, but the self-fields $W^{\pm}$, $Z$ and $A$, although  they are also not the quantum fields in the CGSW model, nevertheless they are the classical counterparts of the Standard Model (SM) bosonic fields and can be named after them. \\
Finally, the question remains as to what is the host object for the  droplet of the bgfms configuration? Let us begin with the similarity of an electron in an atom. The self-field concept, as developed by Barut and Kraus, has been used successfully to compute nonrelativistic and relativistic Lamb shifts
%
%
\cite{bib_B-K-1,relativ_Lamb}.
In their approach, the host object is the electron and the tiny Lamb shift of its wave mechanical energy state arises from the electron fluctuation coupled self consistently to its classical electromagnetic self-field. The self-consistent solution for the Lamb shift is then obtained iteratively (and because of this it is sometimes seen as inferior to the perturbative quantum electrodynamics (QED)).
%
%
In this paper the situation is similar but, the energy of the host fermion (or fermions), if it was, e.g. the electron (or electronic fluctuation), appears to be minute in comparison to the obtained mass of the bgfms configuration.

In Eqs.(\ref{lagrangian})-(\ref{lagrangian fermionic}) the covariant differentiations
$\nabla_{\mu}$ for the scalar fluctuation doublet $\Phi_{f}$ and for a fermionic field fluctuations doublet $L_{f}$ and singlet $R_{f}$ are
\begin{eqnarray}
\label{Pdz_5} \nabla_{\mu}\Phi_{f} = \partial_{\mu}\Phi_{f} + i g W_{\mu} \Phi_{f}
+ \frac{1}{2} i g' Y B_{\mu} \Phi_{f} \;\; ,
\end{eqnarray}
\begin{eqnarray}
\label{Pdz_6}
\nabla_{\mu}L_{f} &=& \partial_{\mu}L_{f} + igW_{\mu}L_{f}
+ \frac{1}{2}ig'YB_{\mu}L_{f} \; , \nonumber \\
\;\;\;\;\; \nabla_{\mu}R_{f} &=& \partial_{\mu}R_{f} +
\frac{1}{2}ig'YB_{\mu}R_{f} \; ,
\end{eqnarray}
where
\begin{eqnarray}
\label{Pdz_8}
W_{\mu} = W_{\mu}^{a} \, \frac{\sigma^{a}}{2}
\end{eqnarray}
is the gauge field decomposition with respect to the $su(2)$ algebra
generators.
The $U_{Y}(1)$ self-field tensor is defined as
\begin{eqnarray}
\label{Pdz_3}
B_{\mu \nu} = \partial_{\mu}B_{\nu} - \partial_{\nu}B_{\mu}
\end{eqnarray}
and the $SU_{L}(2)$ Yang-Mills self-field tensor as
\begin{eqnarray}
\label{Pdz_4}
F_{\mu \nu}^{a} = \partial_{\mu}W_{\nu}^{a} - \partial_{\nu}W_{\mu}^{a} - g\varepsilon_{a b c } W_{\mu}^{b} W^{c}_{\nu} \; ,
\end{eqnarray}
where the symbol $\varepsilon_{a b c}$ signifies the structure constants for
$SU_{L}(2)$, which are
antisymmetric with the interchange of two neighbour indices and $\varepsilon_{1 2 3} = +1 $. \\
The fundamental constants of the model are the coupling constant for $SU_{L}(2)$, which is denoted by $g$,
and the coupling constant for $U_{Y}(1)$, which according to convention is denoted by $g'/2$. The weak hypercharge operator for the $U_{Y}(1)$ group is called $Y$. The quantum numbers in the model are given in the Table (Appendix~1). \\
Now, the scalar fluctuation doublet
\begin{eqnarray}
\label{Pdz_11} \Phi_{f} = \frac{1}{\sqrt {2}} {0 \choose \varphi_{f}}
\end{eqnarray}
contains the scalar field fluctuation $\varphi_{f}$.
We have adopted the notation
\begin{eqnarray}
\label{Pdz_12} L_{f} = {\nu_{f  L} \choose \ell_{f  L}}\;\; {\rm and}
\;\; R_{f} = (\ell_{f  R}) \; ,
\end{eqnarray}
where for the sake of
transparency only one leptonic fluctuation $\ell$ inside the bgfms and its neutrino  fluctuation are specified.
The contribution from other existing fermionic fluctuations can be treated in a similar way.

Now, for our charged (electroweak or weak) physical configuration at
$\varphi_{f} = \delta$, we decompose the {\it total} self-fields
$W^{a}_{\mu}$, $B_{\mu}$ and the scalar field fluctuation $\varphi_{f}$,  which stay on the L{\small HS} of Eq.(\ref{shifts}) as follows
\begin{eqnarray}
\label{shifts}
\left\{ \begin{array}{lll}
W^{a}_{\mu} = \omega^{a}_{\mu} + \tilde{W}^{a}_{\mu} \; , \\
B_{\mu} = b_{\mu} + \tilde{B}_{\mu}  \; ,  \\
\varphi_{f} = \delta + \tilde{\varphi}_{f}  \; .
\end{array} \right.
\end{eqnarray}
Here, each of the total fields on the R{\small HS} is decomposed into the {\it  self consistently} treated  part  $\omega^{a}_{\mu}$, $b_{\mu}$ and $\delta$ and {\it the wavy} (non-self consistent) part $\tilde{W}^{a}_{\mu}$, $\tilde{B}_{\mu}$ of the self-fields and $\tilde{\varphi}_{f}$ of the scalar field fluctuation, respectively. The wavy terms are not treated self consistently. In this paper the thin wall approximation is used in which
$\omega^{a}_{\mu}$, $b_{\mu}$ and $\delta$ are constant.
These homogenous components of the self-fields are the main quantities
which we are interested in and
they are searched for self consistently on the ground
state denoted as $\left( \; \right)_{0}$. The other, wavy parts of the self-fields do not enter into the self-consistent calculation in the presented model.
%
%
Nevertheless, the wavy parts are important in determining
the modified mixing angle $\Theta$ (see Eq.(\ref{Pdz_52})) and in estimating the range of the validity of the thin wall approximation.
%
%
%

\subsection{The screening condition of the fluctuation of charges}

\label{The screening condition}

Now, the effective potential on the ground state is given by
\begin{eqnarray}
\label{Pdz_20}
{\cal U}^{ef}_{f} = - \left( {\cal L}_{f} \right) _{0} \; ,
\end{eqnarray}
where
%
%
${\cal L}_{f}$ is the Lagrangian density (see Eq.(\ref{lagrangian})) of the CGSW model.
Let $J_{f Y}^{\, \mu}$ and $J_{f}^{\, a \mu}$ be the continuous matter current density fluctuations extended in space (see Eqs.(\ref{Pdz_17}-\ref{Pdz_18})) equal on the ground state to
\begin{eqnarray}
\label{Pdz_27}
J_{f Y}^{\, \mu} =  \left( \overline{L_{f}}\gamma^{\mu} Y
L_{f}  +  \overline{R_{f}}\gamma^{\mu} Y R_{f} \right)_{0}   \;\; \nonumber
\\
{\rm
and} \;\;  J_{f}^{\, a \mu} = \left( \overline{L_{f}}\gamma^{\mu}\frac{\sigma^{a}}{2} L_{f} \right)_{0} \;
,
\end{eqnarray}
respectively.
\\
We now assume that on the ground state, for the system in the local rest coordinate system we have
\begin{eqnarray}
\label{Pdz_29}
J_{f Y}^{\, 0} = \varrho_{f Y}\;\; , \;\; J_{f
Y}^{i} = 0\;\; , \;\;J_{f}^{\, a 0} = \varrho_{f}^{\, a}\;\; {\rm
and} \;\;J_{f}^{\, a i} = 0 \; ,
\end{eqnarray}
where $\varrho_{f Y}$ and $\varrho_{f}^{\, a}$ are the matter
charge density fluctuations related to $U_{Y}(1)$ and $SU_{L}(2)$,
respectively.
%
%
%
Eq.(\ref{Pdz_29}) determines
the ground state
which is not relativistically covariant, hence
locally, inside the discussed droplets of the
fluctuations, the Lorentz invariance might not be its fundamental property (the symmetry of the Lagrangian density (\ref{lagrangian fermionic}) still remaining). Yet, we
will see that their diameter in the analyzed cases is
only of the order of $0.001 \; fm$ (see Sections~\ref{mass of charged bgfms} and \ref{mass of neutral bgfms}).\\
{\bf Remark}:
%
%
This means that although some
characteristics of these objects may be detectable, the effects of
the violations of the Lorentz invariance might remain undetectable
or marginally detectable in the present experiments. Similar to the case of partons, which although small are observed, although not
all of their characteristics are detectable. The literature on the
possibility of the violation of the Lorentz invariance is notable
\cite{Lorentz_Ferrero,Lorentz_Diaz,Lorentz_Peck}. \\
\\
As all of the analyses in this paper that pertain to the ground fields are performed on the ground state, therefore, if it is not necessary,  the denotation
$\left(  \, \right)_{0}$ will be omitted.
\\
Thus, what will be finally
found is really the ground state of a system, which follows from the fact that the analyzed droplets of the fields of the excited configurations that lie near the physically interesting solutions have real non-negative squared masses of all their constituent fields. The stability of solutions for the particular configurations of fields is one of the basic problems analyzed in this paper. The particular ground state configurations can decay via radiation or the decay of the
constituent fields only.
There were attempts to approach to such phenomena on the basis of the self-energy rather than on the basis of the quantized radiation field \cite{Barut-Huele-1985}.
\\
%
%
The self-fields which are calculated from (\ref{Pdz_20}) are the ground state fields and only these self-fields are treated fully self consistently in this model.
The boson fields, $W^{a}_{\mu}$, $B_{\mu}$ and
$\varphi_{f}$ (see Eq.(\ref{shifts})), which in the ground state of the whole configuration of fields
are naturally called the ground fields, are denoted as
$\omega^{a}_{\mu}$, $b_{\mu}$ and $\delta$, respectively
\begin{eqnarray}
\label{Pdz_21}
\!\!\!
{\rm self\!\!-\!consistent \; (parts \; of) \; self\!\!-\!fields} \;\; \left\{ \begin{array}{lll}
W^{a}_{\mu} =  \omega^{a}_{\mu} \; , \\
B_{\mu} =  b_{\mu} \; ,  \\
\varphi_{f} =  \delta \; .
\end{array} \right.
\end{eqnarray}
They are searched for self consistently. \\
Next, we assume that also in the decomposition (\ref{shifts}) in the excited states of the system, the self-consistent parts $\omega^{a}_{\mu}$, $b_{\mu}$ of the self-fields and $\delta$ are found from the self-consistent analysis of potential ${\cal U}^{ef}_{f}$ given by Eq.(\ref{Pdz_20})
and that in the excited states matter current density fluctuations are the same as  $J_{f Y}^{\, \mu}$ and $J_{f}^{\, a \mu}$ given by Eqs.(\ref{Pdz_27}),(\ref{Pdz_29}).\\
The self-consistent parts (both on the ground state and on the excited ones) can be parameterized in the following way \cite{JacekManka}
\begin{eqnarray}
\label{Pdz_22}
\omega^{a}_{\mu} =
\left\{\begin{array}{lll}
\omega^{a}_{0} = \sigma \,  n^{a} \; , \\
\omega^{a}_{i} = \vartheta \, \varepsilon_{a i b } \, n^{b} \;\;\;
{\rm and} \;\;\; n^{a}n^{a} = 1 \; ,
\end{array}
\right.
\end{eqnarray}
\nopagebreak[4]
\begin{eqnarray}
\label{Pdz_23}
b_{\mu} =
\left\{\begin{array}{lll}
b_{0} = \beta \; ,\\
b_{i} = 0 \; .
\end{array}
\right.
\end{eqnarray}
In Eq.(\ref{Pdz_22}) the $(n^{a}) = constant$
plays the role of a unit vector in
the adjoint representation of the Lie algebra $su(2)$. It
determines the direction of the ground fields (or more generally of the self-consistent part of the self-fields). It can be seen that
(no summation over index ``a'')
\begin{eqnarray}
\label{Pdz_24}
\!\!\!
\omega^{a}_{\mu}\omega^{a \mu} = \sigma^{2} n^{a} n^{a} -
\vartheta^{2} \varepsilon_{a i b } \varepsilon_{a i b } n^{b} n^{b}  \;\;\;
{\rm and} \; \; \; b_{\mu}b^{\mu} = \beta^{2}  .
\end{eqnarray}
Now, further calculations are performed in the thin wall approximation in which $\omega^{a}_{\mu}$, $b_{\mu}$
and $\delta$ are the homogenous~fields. \\
Using Eqs.(\ref{Pdz_21})-(\ref{Pdz_23}) in Eqs.(\ref{Pdz_20}) and (\ref{lagrangian}), we obtain the effective potential
\begin{eqnarray}
\label{Pdz_30}
{\cal U}^{ef}_{f}(\vartheta,\sigma,\beta,\delta) \!\! & =
& \!\! -g^{2}\sigma^{2}\vartheta^{2} + \frac{1}{2}g^{2}\vartheta^{4} -
\frac{1}{8}g^{2}\delta^{2}(\sigma^{2} - 2 \vartheta^{2})   \nonumber \\
\nonumber & + & \frac{1}{4}gg'\delta^{2}\beta\sigma n^{3} -
\frac{1}{8}g'^{2}\delta^{2}\beta^{2} + g
\varrho_{f}^{\, a}n^{a}\sigma  \nonumber  \\
& + & \frac{g'}{2}\varrho_{f Y}\beta
+ \frac{1}{4}\lambda(\delta^{2} - v^{2})^{2} \; ,
\end{eqnarray}
for the  self-consistent parts of the {\it self-fields}.
For the self-fields on the ground state, the potential ${\cal U}^{ef}_{f}(\vartheta,\sigma,\beta,\delta)$ forms the complete effective potential. \\
When the self-consistent parts of fields are homogenous in time and space, then $\vartheta$, $\sigma$, $\beta$ and $\delta$ are constant
and from $\partial_{\nu} \vartheta = \partial_{\nu} \sigma = \partial_{\nu} \beta = \partial_{\nu} \delta= 0$, $\nu = 0,1,2,3$, it follows that $\nabla^{2} \vartheta = \nabla^{2} \sigma = \nabla^{2} \beta = \nabla^{2} \delta= 0$. This means that (in the thin wall approximation) the self-consistent part of the self-fields and the scalar field fluctuation form an incompressible matter.
%
%
Then, the field equations Eqs.(\ref{Pdz_14})--(\ref{Pdz_16}) and Eq.(\ref{Pdz_19}) (see Appendix~2) that resulted from the CGSW Lagrangian (\ref{lagrangian}) give the following four algebraic equations for the self-consistent parts  $\vartheta$, $\sigma$, $\beta$ of the self-fields and  $\delta$ of the scalar field fluctuation
\begin{eqnarray}
\label{Pdz_32}
\left[ \;\; \frac{1}{2}\delta^{2} - 2 \sigma^{2} + 2 \vartheta^{2} \;\;
\right] \vartheta = 0 \; ,
\end{eqnarray}
\begin{eqnarray}
\label{Pdz_33} - g (2 \vartheta^{2} + \frac{1}{4} \delta^2) \sigma
+ \frac{1}{4} g'\delta^{2} \beta n^{3} + \varrho_{f}^{\, a} n^{a}
= 0 \; ,
\end{eqnarray}
\begin{eqnarray}
\label{Pdz_34} \frac{1}{2} (g\sigma n^{3} - g'\beta)\delta^{2} +
\varrho_{f Y} = 0 \; ,
\end{eqnarray}
\begin{eqnarray}
\label{Pdz_35}
\Biglb[ \; - \, \frac{1}{4}g^{2}(\sigma^{2} - 2 \vartheta^{2}) +
\frac{1}{2}gg'\sigma \beta n^{3} - \frac{1}{4} g'^{2} \beta^{2}
 \nonumber \\
+ \, \lambda \, (\delta^{2} - v^{2}) \; \Bigrb] \; \delta = 0 \; .
\end{eqnarray}
%
%
In the self-consistent homogenous case,  Eqs.(\ref{Pdz_14})--(\ref{Pdz_16}) and Eq.(\ref{Pdz_19}) are equivalent to
\begin{eqnarray}
\label{Pdz_31}
\partial_{\vartheta}{\cal U}^{ef}_{f} = \partial_{\sigma}{\cal U}^{ef}_{f} =
\partial_{\beta}{\cal U}^{ef}_{f} = \partial_{\delta}{\cal U}^{ef}_{f} = 0 \; ,
\end{eqnarray}
and thus Eqs.(\ref{Pdz_32})-(\ref{Pdz_35}) can be easily checked.
They form the self-consistent part of the {\it screening condition of the fluctuation of charges}, which is the analog of the screening current condition in electromagnetism \cite{Aitchison-Hey-bis}. They are used in the calculations of
the value of change of the observed electric and weak density fluctuations of charges (see Eqs.(\ref{Pdz_49})-(\ref{Pdz_51}) below) and the
effective masses of the fields (see Eqs.(\ref{Pdz_37})-(\ref{Pdz_40b}) below).
The self-fields  obtained self consistently, i.e. according to Eqs.(\ref{Pdz_32})-(\ref{Pdz_35}), will be called the {\it self-consistent fields}. The configuration of the self-consistent fields {\it on the ground state}
is called (in agreement with the Introduction) the (boson) {\it
ground fields induced by matter sources} (bgfms) configuration   \cite{JacekManka}. (They can be equivalently obtained self consistently from the effective potential given by Eq.(\ref{Pdz_30}) and Eq.(\ref{Pdz_31}).) \\

When we define the ``electroweak magnetic field'' as ${\cal
B}^{a}_{i} = 1/2 \varepsilon_{i j k } F^{a}_{j k }$ and the
``electroweak electric field'' as ${\cal E}^{a}_{i} = F^{a}_{i 0 }$,
then their self-consistent
parts $(\sigma = constant , \vartheta =
constant , \beta = constant , \; (n^{a}) = constant)$
for $\vartheta \neq 0$
%
%
are equal to
$\left( {\cal B}^{a}_{i} \right)_{0}$ and
%
%
$\left( {\cal E}^{a}_{i} \right)_{0}$, respectively
\cite{JacekManka}
\begin{eqnarray}
\label{Pdz_25}
\!\!\!
\left( {\cal B}^{a}_{i} \right)_{0} = - g \vartheta^{2}n^{i}n^{a} \;\;\; {\rm and}
\;\;\; \left( {\cal E}^{a}_{i} \right)_{0} =
g \sigma \vartheta (\delta_{a i } - n^{a} n^{i}) \, .
\end{eqnarray}
Now, let us choose
\begin{eqnarray}
\label{Pdz_36}
(n^{a}) = (0,0,1) \; .
\end{eqnarray}
In this case
the self-consistent parts of the electroweak magnetic
field $\left( {\cal B}^{3}_{3} \right)_{0} = - g \vartheta^{2}$ along the $x^{3}$ spatial direction and of the electroweak electric
field $\left(  {\cal E}^{1}_{1} \right)_{0} = \left( {\cal E}^{2}_{2} \right)_{0} = g \sigma \vartheta$
pointing in the $x^{1}$ and $x^{2}$ spatial directions,
respectively, are different from zero. \\

Let us perform (for $\delta \neq 0$), a ``rotation'' of  $W^{3}_{\mu}$ and $B_{\mu} $ self-fields to the physical self-fields $Z_{\mu} $ and $ A_{\mu}$
\begin{eqnarray}
\label{Pdz_41}
{Z_{\mu} \choose A_{\mu}} = {{cos\Theta\;\; {-sin\Theta}}
\choose{sin\Theta\;\;\;\;\; cos\Theta}}{W^{3}_{\mu}\choose B_{\mu}} \; .
\end{eqnarray}
Then, consequently a rotation of  $ \sigma $ and $\beta$ self-consistent fields to their counterparts $\zeta$ and $\alpha$ (and similarly for $\tilde{Z}_{\mu} $ and $ \tilde{A}_{\mu}$) as well as a rotation of the charge density fluctuations $\varrho_{f}^{\, 3}$ and $\varrho_{f Y}$ to their corresponding physical quantities $\varrho_{f Z}$ and $\varrho_{f Q}$ are as follows
\begin{eqnarray}
\label{Pdz_42}
{\zeta \choose \alpha} = {{cos\Theta\;\; {-sin\Theta}}\choose{sin\Theta\;
\;\;\;\; cos\Theta}}{\sigma \choose \beta} \; ,
\end{eqnarray}
\begin{eqnarray}
\label{Pdz_43} {(g/cos\Theta) \varrho_{f Z} \choose (g sin\Theta)
\varrho_{f Q}} =
{{cos\Theta\;\;{-sin\Theta}}\choose{sin\Theta\;\;\;\;\;
cos\Theta}}{(g) \varrho_{f}^{\, a} n^{a} \choose (g'/2) \varrho_{f
Y}} \; .
\end{eqnarray}
It is worthwhile to write the relations between weak isotopic
charge density fluctuation $\varrho_{f}^{\, 3}$ (see
Eq.(\ref{Pdz_29}) and Eq.(\ref{Pdz_36})), weak hypercharge
density fluctuation $\varrho_{f Y},$ standard relation (SR) unscreened
electric charge density fluctuation $\varrho_{f Q \;SR}$
(Eq.(\ref{Pdz_51}) below), standard (SR) unscreened weak charge density
fluctuation $\varrho_{f Z \; SR}$ (Eq.(\ref{Pdz_51}) below) and their
generalizations in our model, i.e. the observed electric charge
density fluctuation $\varrho_{f Q}$ and the observed weak charge
density fluctuation $\varrho_{f Z}$
\begin{eqnarray}
\label{Pdz_49}
\varrho_{f Q} = \varrho_{f Q \; SR } + \frac{1}{2}
(\frac{g'}{g} ctg\Theta - 1) \varrho_{f Y} \; ,
\end{eqnarray}
\begin{eqnarray}
\label{Pdz_50}
\varrho_{f Z} = \varrho_{f}^{\, 3} - \varrho_{f Q}
\; sin^{2}\Theta \; ,
\end{eqnarray}
\begin{eqnarray}
\label{Pdz_51}
\varrho_{f Q \; SR} &=& \varrho_{f}^{\, 3} +
\frac{1}{2} \varrho_{f Y} \;\;\;  \nonumber \\
{\rm and} \;\;\; \varrho_{f Z \; SR} &=& \varrho_{f}^{\, 3} - \varrho_{Q \; SR} \; sin^{2}\Theta_{W}
\; .
\end{eqnarray}
Here, $\Theta$ is the modified mixing angle (given below), whereas the  Standard Model (SM) relations between the Weinberg angle
$\Theta_{W}$, $g$ and $g'$ are given by $cos\Theta_{W} = \frac{g}{\sqrt{g^{2} + g'^{2}}}$ and $sin\Theta_{W} = \frac{g'}{\sqrt{g^2 + g'^{2}}}$.\\
\\
The numerical calculations
are performed with the Fermi coupling constant equal to $G_{F} \approx 1.16638 \times 10^{-5} \; {\rm GeV}^{-2}$, the SM value for the boson $W^{\pm}$ mass, $m_{W\,SM} \approx 80.385$ ${\rm GeV}$, and
$sin^{2}\Theta_{W} \approx 0.23116$  \cite{particle_data}.
From these values, $g = \sqrt{8 \, m_{W\,SM}^{2} \, G_{F}/\sqrt{2}\;} \approx 0.65295$ and $g' = g \, tg\Theta_{W} \approx 0.35803$ and $v = 2 \, m_{W\,SM}/g$ $\approx246.22 \; {\rm GeV}$ are
calculated.
The accuracy of the results is restricted by the accuracy of the measurement of the boson $W^{\pm}$ mass $(80.385 \pm  0.015 \; {\rm GeV})$  \cite{particle_data}, i.e., to the fourth significant digits.

\subsection{The masses of the self-fields and scalar field fluctuation}

\label{masses of self and scalar fields}

The massive Lagrangian density for the boson self-fields and the scalar field fluctuation, which follows from the kinematical part of the Lagrangian (\ref{lagrangian}) is equal to
\begin{eqnarray}
\label{lagrangian mass}
{\cal L}_{mass} \!  &=& \!\!
- \frac{1}{2} g^2 \varepsilon_{abc} \varepsilon_{ade} \omega^{b}_{\mu} \omega^{d \, \mu} {\tilde W}^{c}_{\nu} {\tilde W}^{e \, \nu}
+ \frac{1}{8} g^2 \delta^2  {\tilde W}^{a}_{\mu} {\tilde W}^{a \, \mu} \nonumber \\
& - & \frac{1}{4} g g' \delta^2  {\tilde W}^{3}_{\mu} {\tilde B}^{\mu}
+  \frac{1}{8} g'^2 \delta^2  {\tilde B}_{\mu} {\tilde B}^{\mu}
 \\
\!\!\!\!\! \!\!\!\!\! \!\!\!\!\!  \!\!\!\!\! \!\!\!\!\!\!\!\!\!\!
&+&  \frac{1}{8} g^2 \omega^{a}_{\mu} \omega^{a \, \mu} {\tilde \varphi}_{f}^{2} - \frac{1}{4} g g' \omega^{3}_{\mu} b^{\mu}  {\tilde \varphi}_{f}^{2}
+  \frac{1}{8} g'^2 b_{\mu} b^{\mu} {\tilde \varphi}_{f}^{2} \nonumber \\
&-& \frac{1}{2} \lambda (3 \delta^2 - v^{2}) {\tilde \varphi}_{f}^{2} \; . \nonumber
\end{eqnarray}
This changes the effective potential (\ref{Pdz_20}) for the excited states by $\tilde{{\cal U}}^{ef}_{f} = - {\cal L}_{mass}$. \\
Using
Eqs.(\ref{Pdz_21})-(\ref{Pdz_24}) and Eq.(\ref{Pdz_36})
in the massive Lagrangian density (\ref{lagrangian mass}),
we obtain the following squared masses \cite{JacekManka} for (the wavy parts of) the boson self-fields and the scalar field fluctuation (\ref{shifts}) inside a droplet of the bgfms configuration
\begin{eqnarray}
\label{Pdz_37}
m_{{\tilde W}^{1,2}}^{2} = g^{2} (\frac{1}{4}\delta^{2} - \sigma^{2} +
\vartheta^{2}) \; ,
\end{eqnarray}
\begin{eqnarray}
\label{Pdz_38}
m_{{\tilde W}^{3}}^{2} =  g^{2} (\frac{1}{4} \delta^{2} + 2 \vartheta^{2}) \; ,
\end{eqnarray}
\begin{eqnarray}
\label{Pdz_39}
m_{{\tilde B}}^{2} = \frac{1}{4} g'^{2} \delta^{2} \; ,
\end{eqnarray}
%
%
\begin{eqnarray}
\label{Pdz_40b}
%
%
m_{{\tilde \varphi}_{f}}^{2} =   \lambda (3 \delta^{2} - v^{2}) -
\frac{1}{4}g^{2}(\sigma^{2} - 2 \vartheta^{2}) \nonumber \\
+ \frac{1}{2} \, g g' \sigma \beta n^{3} -  \frac{1}{4} g'^{2} \beta^{2} \;  .
\end{eqnarray}
Let us note that the masses in Eqs.(\ref{Pdz_37})-(\ref{Pdz_40b})
%
%
are modified (near the ground state of the droplet) according to
the self-consistent part of the
screening current condition
given by Eqs.(\ref{Pdz_32})-(\ref{Pdz_35}).
\\
After using Eq.(\ref{Pdz_41}), we pass from the fields $\tilde{B}$ and $\tilde{W}^{3}$ to their physical linear combinations $\tilde{A}$ and $\tilde{Z}$ and from (\ref{lagrangian mass}), we obtain their squared masses
\begin{eqnarray}
\label{Pdz_45}
m_{{\tilde Z}}^{2} &=& \frac{1}{2}  \Biglb[ \; m_{Z \;
SR}^{2} + 2 g^{2} \vartheta^{2}  \nonumber \\
&+&  \sqrt{(m_{Z \; SR}^{2} + 2
g^{2} \vartheta^{2})^{2} - 2(gg'\delta \vartheta)^{2}} \;\;
\Bigrb] \; ,
\end{eqnarray}
\begin{eqnarray}
\label{Pdz_46} m_{{\tilde A}}^{2} &=& \frac{1}{2} \Biglb[ \;\; m_{Z \;
SR}^{2} + 2 g^{2} \vartheta^{2}  \nonumber \\
&-&  \sqrt{(m_{Z \; SR}^{2} + 2
g^{2} \vartheta^{2})^{2} - 2(gg'\delta \vartheta)^{2}}  \;\;
\Bigrb] \; ,
\end{eqnarray}
where from the orthogonality property of the mass matrix of the fields $\tilde{A}$ and $\tilde{Z}$, the modified mixing angle $\Theta$ is  obtained
\begin{eqnarray}
\label{Pdz_52}
tg\Theta &=&  \;\; \frac{-
(1 + 8 (\vartheta/\delta)^{2})g^{2} + g'^{2}}{2 g g'}  \nonumber \\
&+&  \sqrt{(
\frac{(1 + 8 (\vartheta/\delta)^{2})g^{2} - g'^{2}}{2 g g'})^{2}
+ 1 \,}  \; .
\end{eqnarray}
In Eqs.(\ref{Pdz_45})-(\ref{Pdz_46}) $m_{Z \; SR}^{2}$ looks similar to the standard relation (SR) for the boson $Z^{\mu}$ squared  mass
\begin{eqnarray}
\label{Pdz_48}
m_{Z \; SR}^{2} \equiv \frac{1}{4}(g^{2} + g'^{2})
\delta^{2} \; .
\end{eqnarray}
Defining the complex self-fields $W^{\pm}_{\mu} = (W^{1}_{\mu} \mp i W^{2}_{\mu})/ \sqrt{2} $ from Eq.(\ref{lagrangian mass}), the squared  masses also follow (compare Eq.(\ref{Pdz_37}))
\begin{eqnarray}
\label{Pdz_44}
m_{\tilde{W}^{\pm}}^{2} = g^{2} \left[ \;\; \frac{1}{4} \delta^{2} - (\zeta
cos\Theta + \alpha sin\Theta)^{2} + \vartheta^{2} \;\; \right] \; .
\end{eqnarray}
Finally, the squared mass of the scalar field fluctuation is equal to
%
%
\begin{eqnarray}
\label{Pdz_47-2}
%
%
m_{\tilde{\varphi}_{f}}^{2} &=&
\lambda (3 \delta^{2} - v^{2}) -
\frac{1}{\delta^{2}}(m_{\tilde{Z}}^{2} \zeta^{2} - m_{\tilde{A}}^{2}
\alpha^{2}) \nonumber \\
&+&
2 g^{2} \left( \frac{1}{\delta^{2}}(\zeta cos\Theta + \alpha sin\Theta)^{2}
+ \frac{1}{4} \right) \vartheta^{2}  \; .
\end{eqnarray}

From Eqs.(\ref{Pdz_32})-(\ref{Pdz_35}) and (\ref{Pdz_42}), we notice that with the simultaneous change of the signs of $\varrho_{f}^{\, 3}$ and $\varrho_{f Y}$,
the signs of $\beta, \sigma$, $\alpha$ and $\zeta$
also change but such physical characteristics as the modified mixing angle $\Theta$ given by Eq.(\ref{Pdz_52}) and the above masses
of the fields inside the bgfms configuration and the mass of the
droplet of the bgfms configuration calculated (further on) using the potential Eq.(\ref{Pdz_30}) remain invariant.
%
%
\\
The calculations below are carried out in the stationary points given
by Eq.(\ref{Pdz_31}) of the effective potential ${\cal U}^{ef}_{f}$ of the self-consistent fields.
It is not difficult to see that
the solutions of Eqs.(\ref{Pdz_32})-(\ref{Pdz_35}) for
the ground fields in these points of the effective potential ${\cal U}^{ef}_{f}$ split into the two cases discussed below, one for the EWbgfms  configuration and the other for the Wbgfms one. \\
It is evident from Eq.(\ref{Pdz_52}) that the transition from the
zero charge density fluctuations to $\varrho_{f}^{\, 3} \neq 0$,
$\varrho_{f Y} \neq 0$ is associated with the non-linear response
of the system.
It can also be noticed that electroweak SM assumptions, which concern the relations between charges, are
formally recovered for $\vartheta = 0$.
Some quantum numbers of the CGSW $SU_{L}(2) \times U_{Y}(1)$ model are given in the Table in Appendix~1.

%
%

\section{The EWbgfms fields configurations with $\varrho_{f Q \; SR } \neq 0$
}

\label{electric}

Now, Eqs.(\ref{Pdz_32})-(\ref{Pdz_35}) can be rewritten as follows:
\begin{eqnarray}
\label{Pdz_53}
\sigma = \frac{1}{2 g \vartheta^{2} } \varrho_{f Q
\; SR } \; ,
\end{eqnarray}
\begin{eqnarray}
\label{Pdz_54}
\beta = \frac{1}{g'}(g \sigma n^{3} + 2
\frac{\varrho_{f Y}}{\delta^{2}}) \; ,
\end{eqnarray}
\begin{eqnarray}
\label{Pdz_55} \vartheta^{6} + \frac{1}{4} \delta^{2}
\vartheta^{4} - \frac{1}{4 g^{2}} \varrho_{f Q \; SR }^{2} = 0 \;
,
\end{eqnarray}
\begin{eqnarray}
\label{Pdz_56} \delta^{6} + (\frac{g^{2}}{2 \lambda} \vartheta^{2}
- v^{2}) \delta^{4} - \frac{1}{\lambda} \varrho_{f Y}^{2} = 0 \; .
\end{eqnarray}
{\bf Note:} From Eq.(\ref{Pdz_55}) we see that the self-consistent field $\vartheta$ is non-zero only if $\varrho_{f Q \; SR} \neq 0$. We  also see that according to Eq.(\ref{Pdz_56}) (compare Eq.(\ref{Pdz_34})), the non-zero value of $\varrho_{f Y}$ implies the non-zero self-consistent field $\delta \neq 0$ of the scalar fluctuation  $\varphi_{f}$.
\\
Now  Eqs.(\ref{Pdz_21})-(\ref{Pdz_23}) with (\ref{Pdz_42}) read
\begin{eqnarray}
\label{Pdz_58}
\left\{ \begin{array}{lll}
W^{\pm}_{0,3} = 0  \; , \;
W^{\pm}_{1} =  \pm i \vartheta/ \sqrt{2} \; ,  \;
W^{\pm}_{2} =  \vartheta / \sqrt{2}  \; , \\
Z_{i} = 0 \; , \;
Z_{0} =  \zeta \; , \; \;
{\rm where} \;
(\zeta = \sigma cos\Theta - \beta sin\Theta)  \; , \\
A_{i} = 0 \; , \; A_{0} = \alpha \; , \;
{\rm where} \;
(\alpha = \sigma sin\Theta + \beta cos\Theta)  \, , \\
\varphi_{f} =  \delta \; .
\end{array} \right.
\end{eqnarray}
Let us note that the relation between the weak hypercharge
quantum number $Y$ and the electric charge quantum number
$Q$ can be written in the form $Q = p \, Y/2$ (for matter fields),
where the corresponding values of $p\,(p\neq~0)$ are given in the
Table in Appendix~1. Then the relation between the weak hypercharge density fluctuation $\varrho_{f Y}$ and the standard electric
charge density fluctuation $\varrho_{f Q \; SR }$ can also be
written in the form
\begin{eqnarray}
\label{Pdz_57}
\varrho_{f Q \; SR } = p \, \frac{\varrho_{f Y}}{2} \;
,
\end{eqnarray}
where different values of $p$ (see Table) represent
different matter fields which can be the sources of charge
density fluctuations. \\
The above-mentioned
screening charge phenomenon now quantified by Eqs. (\ref{Pdz_53})-(\ref{Pdz_56})
is of crucial importance for the characteristics of the  bgfms  configurations  analyzed below.
When the scalar fluctuation field $\varphi_{f}$ together with $W^{\pm}_{1,2}$, $Z_{0}$, $A_{0}$-gauge self-fields with the non-zero self-consistent parts given by Eq.(\ref{Pdz_58})
are present,
then the electroweak magnetic and electric ground fields (\ref{Pdz_25}) penetrate inside the whole spatially extended fermionic fluctuation.
In their presence, the electroweak force generates an ``electroweak screening fluctuation of charges'' in accord with
%
%
Eqs.(\ref{Pdz_53})-(\ref{Pdz_56}) and Eqs.(\ref{Pdz_49})-(\ref{Pdz_51}).
This is connected with the fact that the basic fermionic field
fluctuation
%
%
carries a non-zero charge.
%
%

\subsection{Characteristics of the EWbgfms  configuration
}

\label{Numerical results for rhoQ}

The solutions of Eqs.(\ref{Pdz_53})-(\ref{Pdz_56}) with the condition (\ref{Pdz_57}) were previously discussed in \cite{JacekManka}.
The numerical results of this analysis for the self-consistent parts of fields,  the scalar fluctuation $\delta$ and self-fields $\beta$, $\sigma$, $\vartheta$ and for the physical self-fields $\alpha$ and $\zeta$ (see Eq.(\ref{Pdz_42})) as functions of the electric charge density fluctuation $\varrho_{f Q}$ for $p=2$ are presented in  Figure~1a. One particular value of $\lambda \approx 0.0652$ has been chosen, the choice of which will be argued later on. The plots for different values of $\lambda$ and $p$ can be found in \cite{Dziekuje_Jacek_nova_2}.
\begin{figure}[here]
\begin{center}
\includegraphics[angle=0,width=75mm]{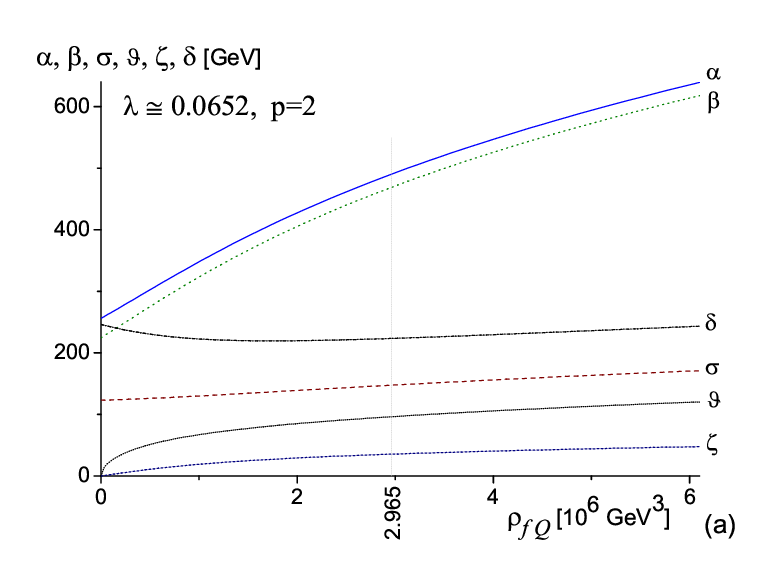}
\includegraphics[angle=0,width=75mm]{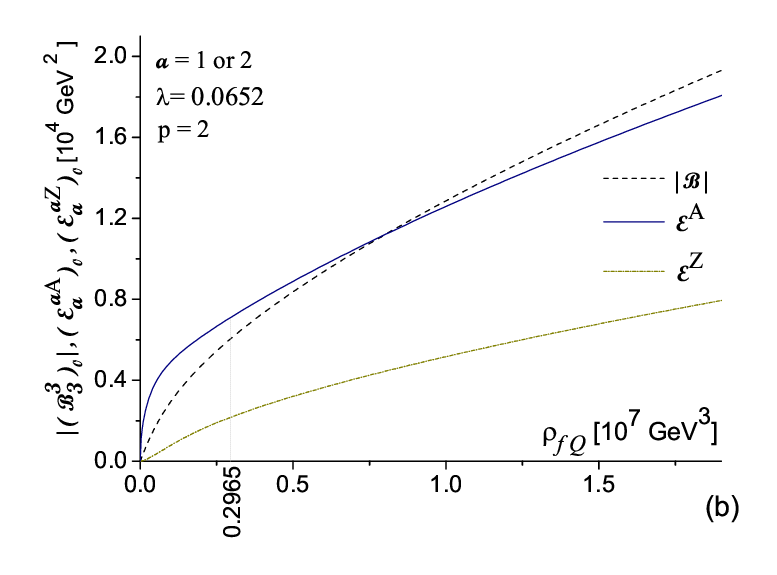}
\end{center}
\caption{{ \it Panel: (a)~The self-consistent parts $\alpha, \beta, \sigma, \vartheta, \zeta$ of the self-fields $A_{0}$, $B_{0}$, $W^{a=3}_{i=0}$, $W^{a=1}_{i=2}$--$W^{a=2}_{i=1}$ and  $Z_{0}$, respectively, as functions of the electric charge density fluctuation $\varrho_{f Q}$($\vartheta \neq 0$, $\delta \neq 0$), Eq.(\ref{Pdz_49}).
The self-consistent field $\delta$ of the self-field $\varphi_{f}$ as the function of
$\varrho_{f Q}$($\vartheta \neq 0$, $\delta \neq 0$).
\newline
(b) The self-consistent parts $\left( {\cal E}^{a\,A}_{a} \right)_{0} = g \, sin\Theta \, \alpha \, \vartheta$,}   (${\cal E}^{A}$), {\it of the ``electromagnetic electric ground fields'' and $\left( {\cal E}^{a\,Z}_{a} \right)_{0} = g \,cos\Theta\, \zeta \, \vartheta$,} (${\cal E}^{Z}$), {\it $a=1,2$ of the ``weak electric ground fields'' (see Eq.(\ref{Pdz_25})-(\ref{Pdz_36}) and Eq.(\ref{Pdz_42})) and  $|\left( {\cal B}^{3}_{3} \right)_{0}| = |- g \vartheta^{2}|$,} ($|{\cal B}|$), {\it of the absolute value of ``electroweak magnetic ground field'' as functions of $\varrho_{f Q}$.
}}
\end{figure}
Here, we notice only that the physical charge density fluctuation
$\varrho_{f Q}$ (see Eq.(\ref{Pdz_49})) for the EWbgfms configuration
for different values of $p$ (see Table)
converge for relatively small values of $\varrho_{f Q}$, i.e. for
values of the charge density fluctuation $\varrho_{f Q}$ in the
range of up to values approximately $10^{3}$ times bigger than those
that correspond to the matter densities in the nucleon.
\begin{figure}[top]
\begin{center}
\includegraphics[angle=0,width=75mm]{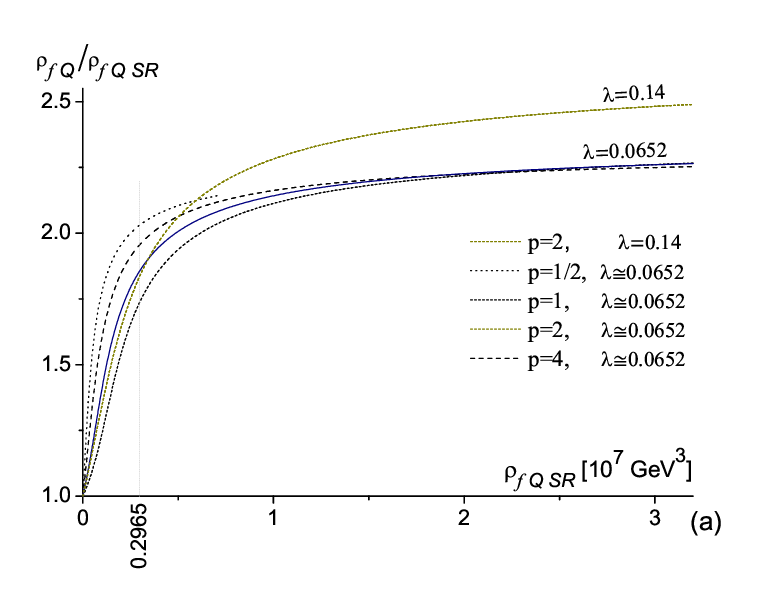}
\includegraphics[angle=0,width=75mm]{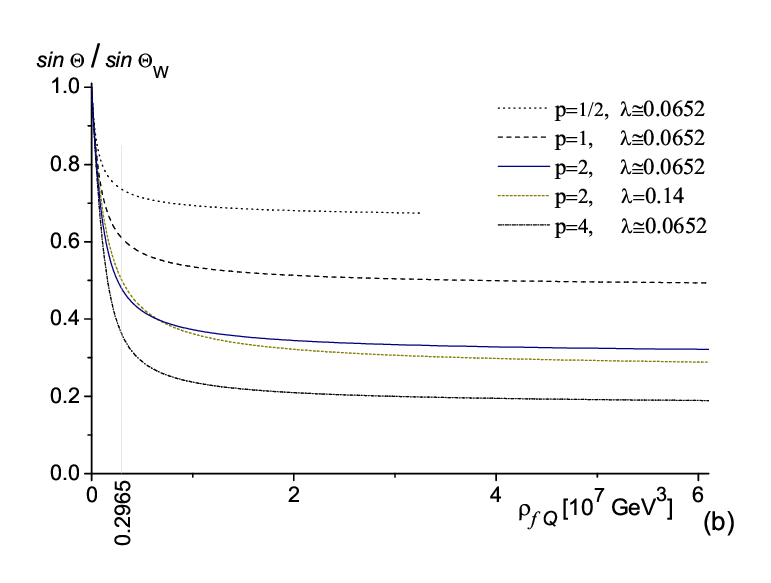}
\caption{ {\it
Panel:~(a) The ratio of the observed electric charge density fluctuation $\varrho_{f Q}$ (see Eq.(44)) to the standard electric charge density fluctuation $\varrho_{f Q \; SR}$($\vartheta \neq 0$,
$\delta \neq 0$) as the function of  $\varrho_{f Q \; SR}$($\vartheta \neq 0$, $\delta \neq 0$).
\newline
(b) The ratio $sin\Theta/sin\Theta_{W}$ (see Eq.(\ref{Pdz_52})) as the function of
$\varrho_{f Q }$($\vartheta \neq 0$, $\delta \neq 0$).
}}
\end{center}
\end{figure}
Also, the ratio  $\varrho_{f Q}/\varrho_{f Q \; SR } \rightarrow 1$ for $\varrho_{f Q \; SR } \rightarrow 0$ (see Figure~2a).
As the result, all of the physical characteristics of the bgfms configurations for different values of $p$ (see Table) converge with $\varrho_{f Q}  \rightarrow 0$ \cite{Dziekuje_Jacek_nova_2}. This can be noticed e.g. from the behavior of the ratio $sin\Theta / sin\Theta_{W}$ (Figure~2b) as a function of $\varrho_{f Q}$,  where
$\Theta$ is the modified mixing angle given by Eq.(\ref{Pdz_52}).
On the other hand, $\varrho_{f Q}/\varrho_{f Q \; SR } \rightarrow C =const > 1$ for $\varrho_{f Q \; SR } \rightarrow \infty$, where the value of $C$ depends both on $p$ and $\lambda$ (see Figure~2a). It can be noticed that the dependance of $C$ on the parameter $\lambda$ of the scalar fluctuation potential is stronger than on $p$. In principle, for bigger values of  $\varrho_{f Q \; SR }$ the information on the true value of $\lambda$ should be extracted from the slope $C$ of the asymptote to the plot of
$\varrho_{f Q}$ as the function of $\varrho_{f Q \; SR }$.
\begin{figure}[here]
\begin{center}
\includegraphics[angle=0,width=75mm]{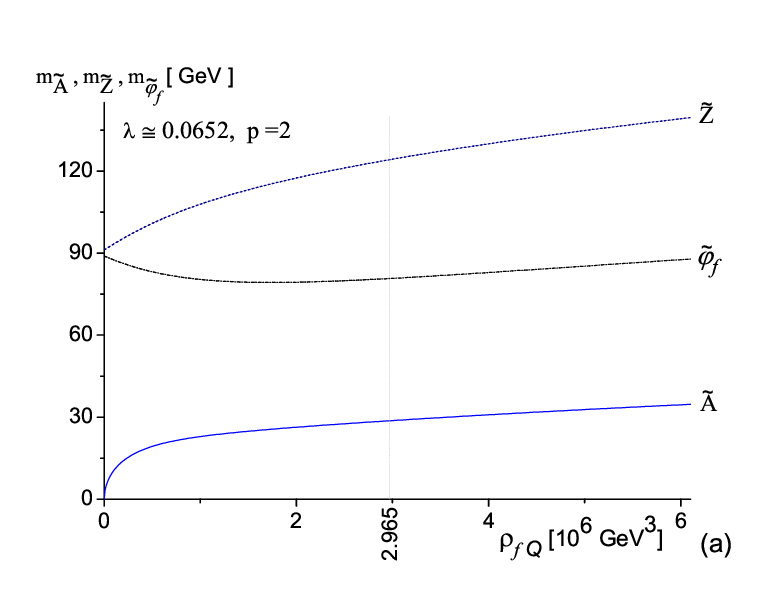}
\includegraphics[angle=0,width=75mm]{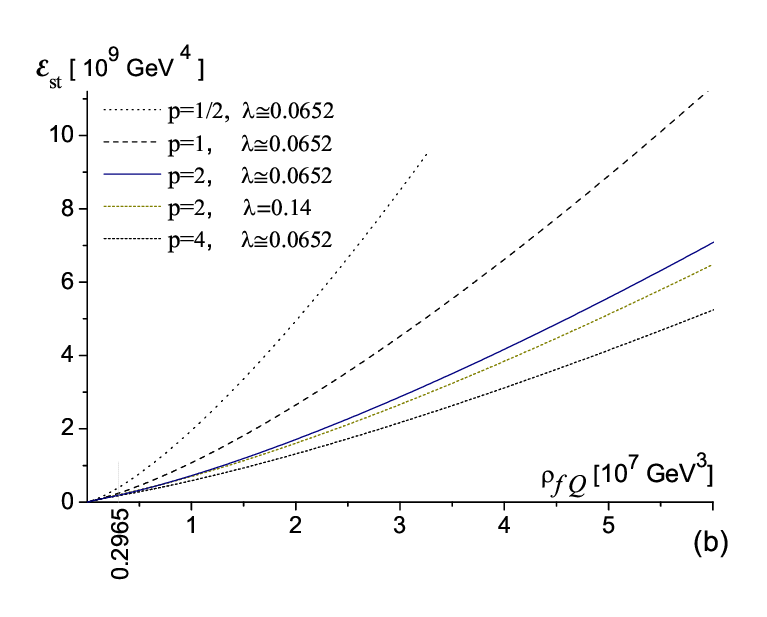}
\caption{ {\it
Panel:~(a) The masses $m_{\tilde{A}}$, $m_{\tilde{Z}}$ and $m_{\tilde{\varphi}_{f}}$ of the gauge boson fields $\tilde{A}^{\mu}$ and $\tilde{Z}^{\mu}$ and of the scalar field fluctuation $\tilde{\varphi}_{f}$, respectively, as functions of the electric charge density fluctuation $\varrho_{f Q }$($\vartheta \neq 0$, $\delta \neq 0$).
\newline
(b) The energy density ${\cal E}_{st}(\varrho_{f Q})$, (\ref{E min rhoQ}),
of the EWbgfms configuration for boson ground fields calculated self consistently according to Eqs.(\ref{Pdz_53})-(\ref{Pdz_56}) (for all
values of $p \neq 0$ from the Table) as the function of
$\varrho_{f Q}$.
}}
\end{center}
\end{figure}\\
From Eq.(\ref{Pdz_32}) and Eq.(\ref{Pdz_37}) (for $\vartheta \neq 0$),
it can be noticed that fields $\tilde{W}^{+}$ and $\tilde{W}^{-}$ (see Eq.(\ref{Pdz_58})), taken together as a pair of massive fields, become {\it inside} the EWbgfms configuration the
{\it massless} self-fields
that are coupled to the charge density
fluctuations $\varrho_{f Q} \neq 0$ ($\varrho_{f Q \; SR} \neq 0$
and $\varrho_{f Y} \neq 0$).
The results for the dependance of the masses of $\tilde{A}$, $\tilde{Z}$ and $\tilde{\varphi}_{f}$ fields (see Eqs.(\ref{Pdz_46})~(\ref{Pdz_45}) and (\ref{Pdz_47-2})) inside the EWbgfms  configuration on
the electric charge density fluctuation $\varrho_{f Q}$($\vartheta \neq 0$, $\delta \neq 0$) are presented in Figure~3a.
%

Let us notice that the expressions (\ref{Pdz_45}) for  $m_{\tilde{Z}}^2$ and  (\ref{Pdz_46}) for $m_{\tilde{A}}^2$  have a root.
For a particular value of $p < 1.388  \approx 2
\sqrt{\sin \Theta_{W}}$
%
%
and below some value of $\lambda = \lambda_{Z}$ (which depends on $p$), the expression $(m_{Z \; SR}^{2} + 2 g^{2} \vartheta^{2})^{2} - 2(gg'\delta \vartheta)^{2}$ under this root gets  above some value of $\varrho_{f Q}$ the negative sign
%
%
so that the EWbgfms configuration becomes unstable in the $\tilde{Z}$ and $\tilde{A}$ field sectors.
Thus, for $p < 1.388$  and a particular value $\lambda < \lambda_{Z}$,  there is a value of $\varrho_{f Q}$ for which $m_{\tilde{A}} = m_{\tilde{Z}}$. \\
For example, for $p=1/2$ the limiting value $\lambda_{Z} \approx 0.2148$. Thus, e.g. for $\lambda = 0.14 < \lambda_{Z}$ this expression becomes negative above $\varrho_{f Q} \approx
1.767 \times 10^{8}$ ${\rm GeV}^{3}$ (for which ${\cal E}_{st} (\varrho_{f Q}) \approx 8.313 \times  10^{10} $ ${\rm GeV}^{4}$).
%
%
For $p=1/2$ and $\lambda =  0.0652 < \lambda_{Z}$ this expression becomes negative above $\varrho_{f Q} \approx
1.531 \times 10^{7}$ ${\rm GeV}^{3}$ (for which ${\cal E}_{st} (\varrho_{f Q}) \approx 3.456 \times  10^{9} $ ${\rm GeV}^{4}$).
%
%
%
Next, e.g. for $p=1$ the limiting value $\lambda_{Z} \approx 0.0297$.
It will be shown in Section~\ref{intersection} that the value of $\varrho_{f Q}$ for a physically interesting EWbgfms configuration (e.g. the state $s2$ in Section~\ref{intersection}) (for which this instability might potentially appear) is smaller than the mentioned limiting value of $\varrho_{f Q}$.
Moreover, above $p \approx 1.388$, and thus also from $p = 3/2$ upwards,  the discussed
%
%
configurations do not possess this instability in the $\tilde{Z}$ and $\tilde{A}$ field sectors for all values of $\lambda$ and $\varrho_{f Q}$.

%
%

\subsection{The mass of the EWbgfms
configuration}

\label{mass of charged bgfms}

The energy density given by Eq.(\ref{Pdz_30}) for stationary ({\it st}) solutions of the EWbgfms configuration for boson ground fields calculated self consistently according to Eqs.(\ref{Pdz_53})-(\ref{Pdz_56}) as the function of $\varrho_{f Q}$ is equal to
\begin{eqnarray}
\label{E min rhoQ}
\!\!\!\!\!\!\!\!\!\! {\cal E}_{st} (\varrho_{f Q}) &=& {\cal U}^{ef}_{f}(\vartheta \neq
0, \delta \neq 0) \nonumber \\
& & ({\rm with \; fields \; treated \; self \; consistently}) \; .
\end{eqnarray}
The energy density ${\cal E}_{st} (\varrho_{f Q})$
increases both with $\varrho_{f Q}$ and $\varrho_{f Q \; SR}$.
The plots of the dependance of ${\cal E}_{st}(\varrho_{f Q})$ for boson ground fields given by Eqs.(\ref{Pdz_53})-(\ref{Pdz_56}) on the electric charge density fluctuation $\varrho_{f Q}$ are presented in Figure~3b (for values of $p \neq 0$ from the Table). We notice
%
%
that from the point of view of ${\cal E}_{st}(\varrho_{f Q})$, the EWbgfms configurations fall into classes of $p$ that differ weakly with $\lambda$  inside a particular class (which is shown in Figure~3b for $p=2$ only).
\begin{figure}[here]
\begin{center}
\includegraphics[angle=0,width=75mm]{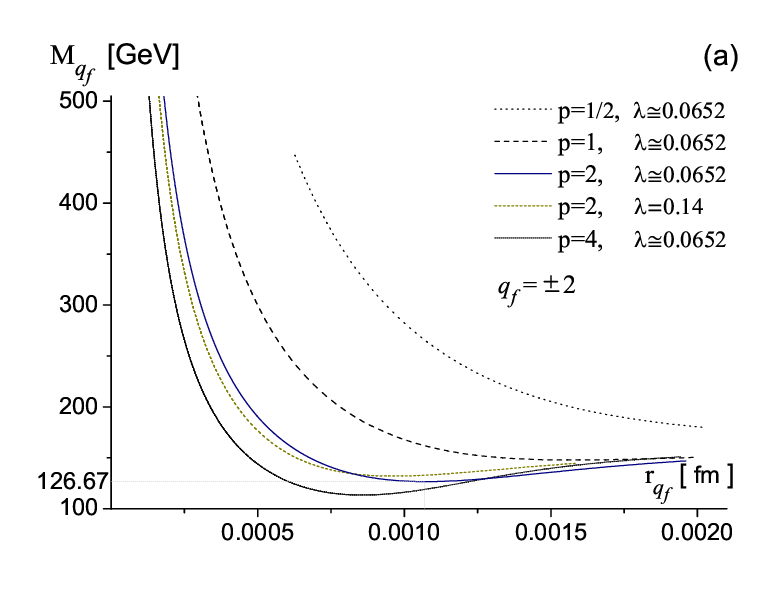}
\includegraphics[angle=0,width=75mm]{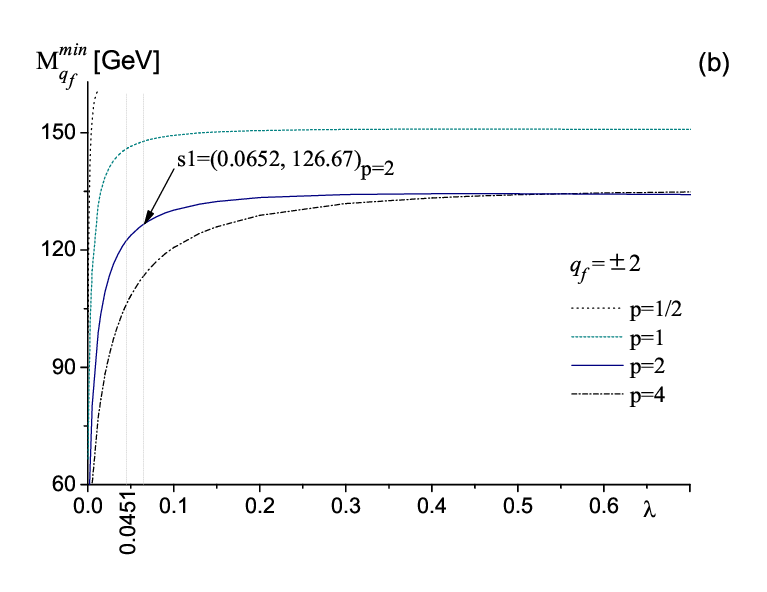}
\caption{
{\it Panel:~(a)
The mass $M_{q_{f} = \pm 2}$ of the EWbgfms configuration as a function of the radius $r_{q_{f}}$ (for
$p \neq 0$ from the Table and exemplary $\lambda$'s). The curves with
$p \geq 1$ exhibit local minima. For example the minimal $M_{q_{f}=\pm 2}$ for $p=2$ and $\lambda \approx 0.0652$ is equal to $M^{min}_{q_{f}=\pm 2}=126.67$ GeV for $r_{q_{f}} \approx 0.00107 \; fm$. The figures are plotted up to the values of $r_{q_{f}}$ smaller than $1/m_{\tilde{Z}}$ (see Figure~3a).
For $p=1/2$ values of $\varrho_{f Q}$ are no
bigger than
$3.297 \times 10^7$ ${\rm GeV^3}$ (for which $r_{q_{f}=2} \approx $ $0.000481 \; fm$) as above it
the configuration becomes unstable ($m_{\tilde{Z}}^{2}$ (\ref{Pdz_45}) and $m_{\tilde{A}}^{2}$ (\ref{Pdz_46}) become imaginary). For $p=1/2$ and  for $\lambda > 0.0119$ there are no EWbgfms configurations with local minimum of $M_{q_{f}}(r_{q_{f}})$.
%
%
\newline
(b) The minimal mass $M^{min}_{q_{f}}$ of the EWbgfms configuration as a function of $\lambda$.
In the case of $p=1/2$,
%
%
the thin wall approximation is not fulfilled
and there are also no EWbgfms configurations with local minimum of $M_{q_{f}}(r_{q_{f}})$ for $\lambda > 0.0119$; hence, we see the cut in the curve  above this value
(compare text under Figure~4a).
}}
\end{center}
\end{figure}

The matter electric charge fluctuation of an electrically charged EWbgfms configuration is equal to
\begin{eqnarray}
\label{qf dla MQ}
q_{f} = \frac{4}{3} \, \pi \, r_{q_{f}}^{3} \, \varrho_{f Q} \; ,
\end{eqnarray}
where $r_{q_{f}}$ is the ``radius of the charge density fluctuation'' in the thin wall approximation. The radius $r_{q_{f}}$ is the function of $\varrho_{f Q}$.
The mass of the electrically charged EWbgfms configuration is equal to
\begin{eqnarray}
\label{M od rQ}
\!\! M_{q_{f}} = \frac{4}{3} \, \pi \, r_{q_{f}}^{3} \; {\cal
E}_{st}(\varrho_{f Q}) \;\;\;\; {\rm and} \;\;\;\; M_{q_{f}} = \pm q_{f} M_{q_{f}=1} \; ,
\end{eqnarray}
where because of the Pauli exclusion principle used for the fermionic fluctuations, we obtain that $q_{f}=\pm 1$ or $\pm 2$ only inside one droplet (except the cases that the consecutive fermionic fluctuations occupy their higher energy states).
%
%
When the fermionic fluctuation (one or two in each bgfms configuration of fields) that plays the role of the matter source that induces boson ground fields was taken into account in the calculation of mass $M_{q_{f}}$, then its value would be changed by an order of the energy of this fermionic fluctuation.
In this paper the energy of the fermionic fluctuation
is neglected. \\
The functional dependence of the mass $M_{q_{f}}$ of a droplet of the EWbgfms configuration of fields (with charge fluctuation $q_{f}$) on $r_{q_{f}}(\varrho_{f Q})$ is presented in Figure~4a.
%
%
It exhibits a minimum in $r_{q_{f}}$ (and also in $\varrho_{f Q}$) for some values of $p$. For instance (see \cite{Dziekuje_Jacek_nova_2}),   for $p = 2$ and with $\lambda \approx 0.0652$, it has the minimal value $M_{q_{f}} = \pm q_{f} \times 63.335 \; GeV$
at $\varrho_{f Q} = 2.965 \times 10^{6} \, {\rm GeV}^{3}$ ($\varrho_{f Q \; SR} = 1.788  \times 10^{6}  \, {\rm GeV}^{3}$) and ${\cal E}_{st}(\varrho_{f Q}) = 1.878 \times 10^{8}  \, {\rm GeV}^{4}$ with the radius of the corresponding charge density fluctuation $r_{q_{f}} = q_{f}^{1/3} \times 0.000852 \; fm$.
In comparison, for a proton with a global electric charge $Q =
1$, its electric charge radius $r_{Q} \approx 0.805 \; fm$.
\\
Finally, let us suppose that in the process, a droplet of the EWbgfms configuration with a particular $p \geq 1$ appears.
This self-consistent charged EWbgfms configuration lies
%
%
in the {\it minimum} of the function of mass $M_{q_{f}=2}$ v.s $\varrho_{f Q}$ (or $r_{q_{f}}$) (see  Figure~4a).
Its self-consistent (homogenous) self-fields are the solution of the equations of motion (\ref{Pdz_14})--(\ref{Pdz_16}) and (\ref{Pdz_19}).  %
%
If necessary, we will mark this {\it minimal mass} by $M^{min}_{q_{f}}$.
This stationary state is the resonance via the weak interactions only,  and can disintegrate through simultaneous decay or radiation of its constituent fields.
%
%
%
The most interesting fact is that the closest configuration of fields is an electrically neutral Wbgfms configuration with the same mass.
%
Because their masses are equal, hence their Breit-Weisskopf-Wigner probability density has a dispersion of the same order. \\
%
%
%
%

{\bf Note:}
From Figure~3b we see that
${\cal E}_{st} \longrightarrow 0 $ as $ \varrho_{f Q \; SR}
\longrightarrow~0$ ($\varrho_{f Q} \longrightarrow~0$) for all of the values of $\lambda > 0$ and $p \neq 0$
considered  (see Table).
%
%
For $ \varrho_{f Q} \longrightarrow 0$ and for all of the considered values of $\lambda >0$ and $p \neq 0$ (see Table) from Eq.(\ref{Pdz_30}) and Eqs.(\ref{Pdz_53})-(\ref{Pdz_56}), we also obtain \begin{eqnarray}
\label{W particle}
M_{q_{f}}
\longrightarrow \pm \, q_{f} \, g \, v/2 = \pm \, q_{f} \times 80.385 \; {\rm  GeV} \; ,
\end{eqnarray}
where the sign ``+'' is for $q_{f} > 0$ and sign ``-'' for $q_{f} < 0$.
Yet, as in this limit the EWbgfms configuration inside a droplet does not reproduce the uncharged SM configuration (for which $ \varrho_{f Q}=0$), thus even for $q_{f} = \pm 1$ this bgfms configuration cannot be interpreted as the observed, well-known $W^{\pm}$ boson particle.
%
%
Indeed, even if the charge density fluctuation tends in the limit to zero  $ \varrho_{f Q} \longrightarrow~0$ and thus we obtain $\vartheta \rightarrow 0$ and $\zeta \rightarrow 0$ for the ground fields of the $W^{+}-W^{-}$ pair and $Z$, respectively, yet, the result is that the self-consistent ground field $\alpha$ of $A_{0}$ is still non-zero in this limit (see Eq.(\ref{Pdz_58}) and Figure~1a)
\cite{Dziekuje_Jacek_nova_2}.
%
%
Therefore, the transition from the configuration of fields with $\varrho_{f Q} \neq 0$ ($\varrho_{f Q  \;
SR} \neq 0$ and $\varrho_{f Y} \neq 0$) to the configuration with
$\varrho_{f Q} = 0$ (then with $\varrho_{f Q  \;
SR} = 0$, $\varrho_{f Y} = 0$, $\alpha=0$, $\zeta = 0$ and $\vartheta = 0$) inside the droplet of the EWbgfms configuration is not a continuous one. Let us notice that
in the double limit $\varrho_{f Q} \longrightarrow 0$ and $q_{f} \longrightarrow 0$, we obtain $M_{q_{f}} = 0$.
%
%

\section{Wbgfms configurations with $\varrho_{f Z \; SR} \neq 0$
}

\label{neutral}

From Eq.(\ref{Pdz_52}) it can be noticed that for $\vartheta  = 0$ the standard relation
$tg \Theta = tg \Theta_{W} = \frac{g'}{g}$ is held; hence, from Eqs.(\ref{Pdz_49})-(\ref{Pdz_51}) it follows that $\varrho_{f Z} = \varrho_{f Z \; SR}$ and $\varrho_{f Q} = \varrho_{f Q \; SR}$.
(The other possibility $tg \Theta = - \frac{g}{g'} = - ctg \Theta_{W}$ obtained in this case from Eq.(\ref{Pdz_52}) is not a physical solution.)\\
Using Eqs.(\ref{Pdz_42})-(\ref{Pdz_43}) we can rewrite the effective potential ${\cal U}^{ef}_{f}$ given by Eq.(\ref{Pdz_30}) for the ground fields in the following form
\begin{eqnarray}
\label{Pdz_67}
\!
{\cal U}^{ef}_{f}(\zeta,\alpha,\delta) \!\!
& = &  \!\! \sqrt{g^{2} + g'^{2}} \, \varrho_{f Z \; SR} \, \zeta + \frac{g\,g'}{\sqrt{g^{2} + g'^{2}}}\, \varrho_{f Q \; SR} \, \alpha \nonumber \\
\!\!
&-& \!\!
\frac{1}{8}(g^{2} + g'^{2})\, \delta^{2}
\zeta^{2} + \frac{1}{4} \lambda \,(\delta^{2} - v^{2})^{2} \, .
\end{eqnarray}
For $\vartheta  = 0$ we can rewrite Eqs.(\ref{Pdz_33})-(\ref{Pdz_35}) as follows
\begin{eqnarray}
\label{Pdz_68}
\varrho_{f Q \; SR} = 0 \; ,
\end{eqnarray}
\begin{eqnarray}
\label{Pdz_69}
\frac{1}{4} \sqrt{g^{2} + g'^{2}} \, \delta^{2}\, \zeta
= \varrho_{f Z \; SR}
\end{eqnarray}
and
\begin{eqnarray}
\label{Pdz_70}
\lambda \, (\delta^{2} - v^{2}) - \frac{1}{4}(g^{2}
+ g'^{2}) \, \zeta^{2} = 0 \; .
\end{eqnarray}
The relations (\ref{Pdz_68})-(\ref{Pdz_70}) form the self-consistent part of the screening condition of the fluctuation of charges. \\
{\bf Note:} Thus, according to Eq.(\ref{Pdz_69}), the non-zero weak charge density fluctuation $\varrho_{f Z \; SR}$
inevitably leads to the non-zero self-consistent field $\zeta$ of $Z_{\mu}$. The non-zero $\varrho_{f Z \; SR}$ also implies the non-zero self-consistent field $\delta \neq 0$ of the scalar fluctuation  $\varphi_{f}$ (compare the Note below Eq.(\ref{Pdz_56})).\\
%
%
Using Eq.(\ref{Pdz_67}) and equations (compare Eq.(\ref{Pdz_31}))
\begin{eqnarray}
\label{stat U dla Z alfa}
\partial_{\alpha}{\cal U}^{ef}_{f} = 0 \; ,
\end{eqnarray}
and
\begin{eqnarray}
\label{stat U dla Z}
\partial_{\zeta}{\cal U}^{ef}_{f} = \partial_{\delta}{\cal U}^{ef}_{f} = 0 \; ,
\end{eqnarray}
the relations (\ref{Pdz_68})-(\ref{Pdz_70}) can easily be checked. \\
Two nontrivial relations given by Eqs.(\ref{Pdz_69}) -- (\ref{Pdz_70}) lead to the solution
\begin{eqnarray}
\label{Pdz_73} \delta^{2}(\varrho_{f Z \; SR}) = \frac{4 \,
\varrho_{f Z \; SR}}{\sqrt{g^{2} + g'^{2}} \, \zeta} \; ,
\end{eqnarray}
and
\begin{eqnarray}
\label{Pdz_72}
\!\!\!\!\!\!\!\!\!\!\!\!\!\!
& & \zeta(\varrho_{f Z \; SR}) = \frac{2}{3^{\frac{1}{2}}\,{\left( g^2 + g'^{\,2} \right) }^{\frac{1}{2}}} \nonumber \\
\!\!\!\!\!\!\!\!\!\!\!\! &\times&
\frac{
     \,\lambda^{- \, \frac{1}{3}} {\left( 3^{\frac{3}{2}}\,\varrho_{f Z \; SR} + {\sqrt{ \, 27\,{\varrho^{\,2}_{f Z \; SR}} + \lambda\,{v}^6 \, }} \,\right) }^
        {\frac{2}{3}} - {v}^2  }{\lambda^{- \, \frac{2}{3}}\,{\left( 3^{\frac{3}{2}}\,\varrho_{f Z \; SR} +
        {\sqrt{ \, 27\,{\varrho^{\,2}_{f Z \; SR}} + \lambda\,{v}^6 \, }} \,\right) }^{\frac{1}{3}}} \; ,
\end{eqnarray}
where self-consistent fields $\zeta$ and $\delta$ are the functions of $\varrho_{f Z \; SR}$ only  (see Figure~5a).
\begin{figure}[here]
\begin{center}
\includegraphics[angle=0,width=75mm]{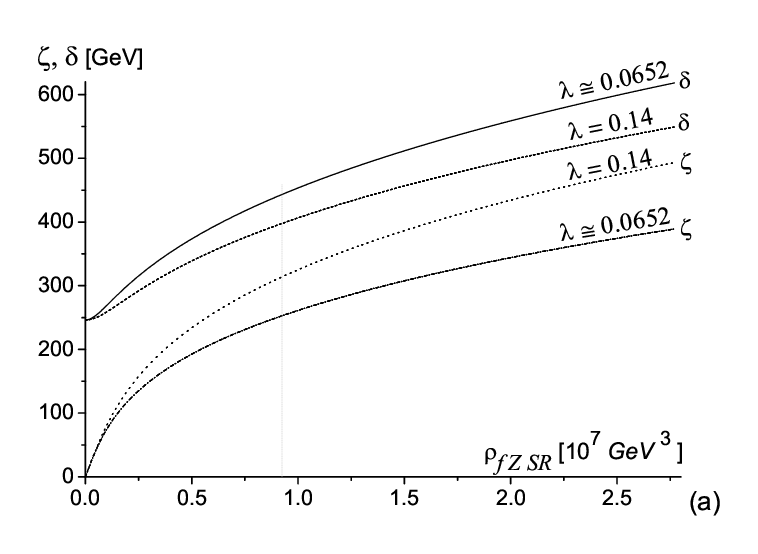}
\includegraphics[angle=0,width=75mm]{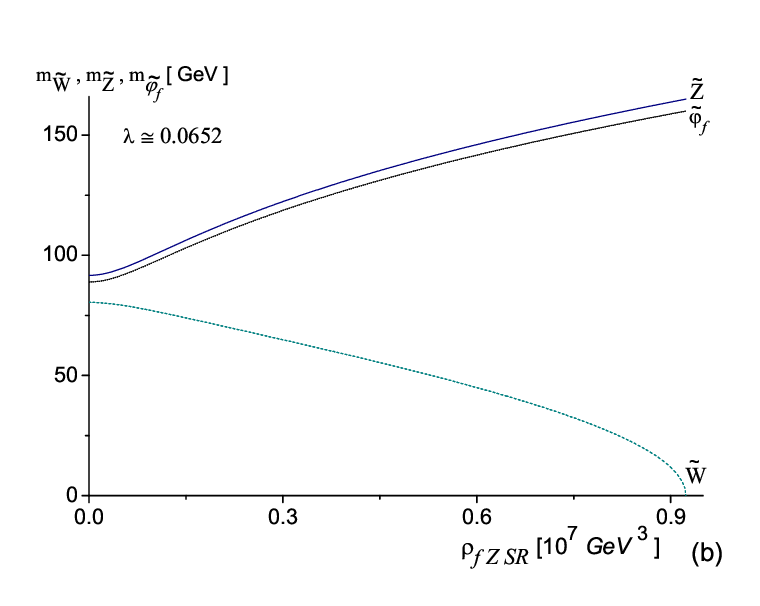}
\end{center}
\caption{{ \it Panel: (a) The self-consistent fields $\zeta$ of $Z_{0}$ and $\delta$ of $\varphi_{f}$ as the function
of the (standard) weak charge density fluctuation $\varrho_{f Z \;
SR}(\vartheta =0,\delta \neq 0, \varrho_{f Q \; SR}=0$).
In the limit $\varrho_{f Z \; SR} \longrightarrow 0$ the self-consistent ground fields tend to the uncharged
values $\zeta=0$ and $\delta=v$.
\newline
(b) The masses $m_{\tilde{W}^{\pm}}$ (Eq.(\ref{Pdz_78})), $m_{\tilde{Z}}$ (Eq.(\ref{Pdz_79})) of the wavy part of $W^{\pm}_{\mu}$ and $Z_{\mu}$, respectively, and the mass $m_{\tilde{\varphi}_{f}}$ (Eq.(\ref{Pdz_81b})) of the wavy part of $\varphi_{f}$
inside the droplet of the Wbgfms configuration as the function of
$\varrho_{f Z \; SR}(\vartheta =0,\delta \neq 0, \varrho_{f Q \; SR}=0$).
In the limit $\varrho_{f Z \; SR} \longrightarrow 0$, these masses tend to the uncharged (i.e. for $\varrho_{f Z \; SR} = 0$) values $m_{W^{\pm}} = g \, v/2$, $m_{Z} = \sqrt{g^2 + g'^2} \, v/2$ and $m_{\varphi_{f}} = \sqrt{2 \lambda} \, v$, respectively.
}}
\end{figure}
Using Eqs.(\ref{Pdz_22})-(\ref{Pdz_23}) and Eqs.(\ref{Pdz_41})-(\ref{Pdz_42}),
we can rewrite Eq.(\ref{Pdz_21}) for the self-consistent field $\alpha$ of  $A_{\mu}$ in the form
\begin{eqnarray}
\label{Pdz_74}
A_{\mu} = (\alpha,0,0,0) \; .
\end{eqnarray}
From Eqs.(\ref{Pdz_68})-(\ref{Pdz_70}) and (\ref{Pdz_72})-(\ref{Pdz_71}) below it follows that
$\alpha$ is not a dynamical variable. It corresponds to a nonphysical degree of freedom and can be removed by the gauge
transformation~$\alpha \rightarrow 0$.
Thus, $U_{Q}(1)$ remains the valid symmetry group giving (see Eq.(\ref{Pdz_42}))
\begin{eqnarray}
\label{Pdz_75}
\alpha = \sigma sin\Theta_{W} + \beta cos\Theta_{W} = 0 \; .
\end{eqnarray}
Now, the self-consistent fields (\ref{Pdz_21})
can be rewritten as follows:
\begin{eqnarray}
\label{Pdz_76}
\left\{ \begin{array}{lll}
W^{1,2}_{\mu} = 0 \;\; , \;\; W^{3}_{i} = 0 \; , \\
W^{3}_{0} = - \beta \, ctg\Theta_{W}  \; , \\
B_{0} = \beta  \; , \\
B_{i} = 0 \; , \\
\varphi_{f} = \delta \; .
\end{array} \right.
\end{eqnarray}
or in terms of physical fields
\begin{eqnarray}
\label{Pdz_77}
\left\{ \begin{array}{lll}
W^{\pm}_{\mu} = 0 \;\; , \;\; Z_{i} = 0 \; ,
\\
Z_{0} = \zeta \;\; , \;\;\;  {\rm where} \;\;\;
(\zeta = -\frac{1}{sin\Theta_{W}} \; \beta)  \; , \\
A_{\mu} = 0  \; , \\
\varphi_{f} = \delta \; .
\end{array} \right.
\end{eqnarray}
The appearance of the non-zero weak charge density
fluctuation $\varrho_{f Z \; SR}$ and the self-consistent field $\zeta$ of the
self-field $Z_{\mu}$ that is induced by it (see Eq.(\ref{Pdz_72})) influences the masses of the wavy parts of the boson self-fields and of the scalar field fluctuation. Their squares inside
a droplet of the Wbgfms configuration are, according to
Eqs.(\ref{Pdz_45})-(\ref{Pdz_46}), (\ref{Pdz_44})-(\ref{Pdz_47-2})
(for $\vartheta = 0$),
equal to  (see Figure~5b)
\begin{eqnarray}
\label{Pdz_79} m_{\tilde{Z}}^{2} = \frac{1}{4}(g^{2} + g'^{2})\delta^{2}
\; ,
\end{eqnarray}
\begin{eqnarray}
\label{Pdz_80}
m_{\tilde{A}}^{2} = 0 \; ,
\end{eqnarray}
%
%
\begin{eqnarray}
\label{Pdz_78}
m_{\tilde{W}^{\pm}}^{2} = \frac{1}{4} g^{2} (\delta^{2} -  4 \,\zeta^{2}
cos^{2} \Theta_{W}) \; ,
\end{eqnarray}
\begin{eqnarray}
\label{Pdz_81b}
%
m_{\tilde{\varphi}_{f}}^{2} =  \lambda \, (3\delta^{2} - v^{2}) - \frac{1}{4}(g^{2} +
g'^{2}) \, \zeta^{2}  \; .
\end{eqnarray}
Thus the effective mass of the wavy part of the physical self-field $A_{\mu}$ is equal to $m_{\tilde{A}} = 0$.\\
\\
After putting the self-consistent ground fields
calculated according to Eqs.(\ref{Pdz_69})-(\ref{Pdz_70}) together with Eq.(\ref{Pdz_68}) into
Eq.(\ref{Pdz_67}) the energy density for the stationary solution of the Wbgfms configuration, ${\cal E}_{st}(\delta, \varrho_{f Z \; SR}) = {\cal U}^{ef}_{f}(\delta, \;
\varrho_{f Z \; SR} ; \; \vartheta = 0,\; \varrho_{f Q \; SR}=0)$ is  obtained \cite{Dziekuje_Jacek_nova_2}
\begin{eqnarray}
\label{Pdz_71}
{\cal E}_{st}(\delta, \varrho_{f Z \; SR}) = 2 \,\frac{\varrho_{f Z SR}^{2}}{\delta^{2}}
+ \frac{1}{4} \, \lambda \, (\delta^{2} - v^{2})^{2}  \;\;\;\;
\end{eqnarray}
%
%
%
(with $\delta$ treated self consistently), which after using Eqs.(\ref{Pdz_69}) and (\ref{Pdz_70}) could also be rewritten as follows (see Figure~6)
\begin{eqnarray}
\label{Pdz_82}
& & {\cal E}_{st}(\varrho_{f Z \; SR})  = \frac{1}{2} \sqrt{g^{2} + g'^{2}} \; \zeta(\varrho_{f Z \; SR}) \nonumber \\
&\times&
\,
\, \left( \varrho_{f Z \; SR} + \frac{1}{32 \, \lambda}(g^{2} + g'^{2})^{\frac{3}{2}} \, \zeta^{3}(\varrho_{f Z \; SR}) \right) \, ,
\end{eqnarray}
where the self-consistent ground field $\zeta = \zeta(\varrho_{f Z \; SR})$ is the function of $\varrho_{f Z \; SR}$ (see Eq.(\ref{Pdz_72})).
\begin{figure}[here]
\begin{center}
\includegraphics[angle=0,width=70mm]{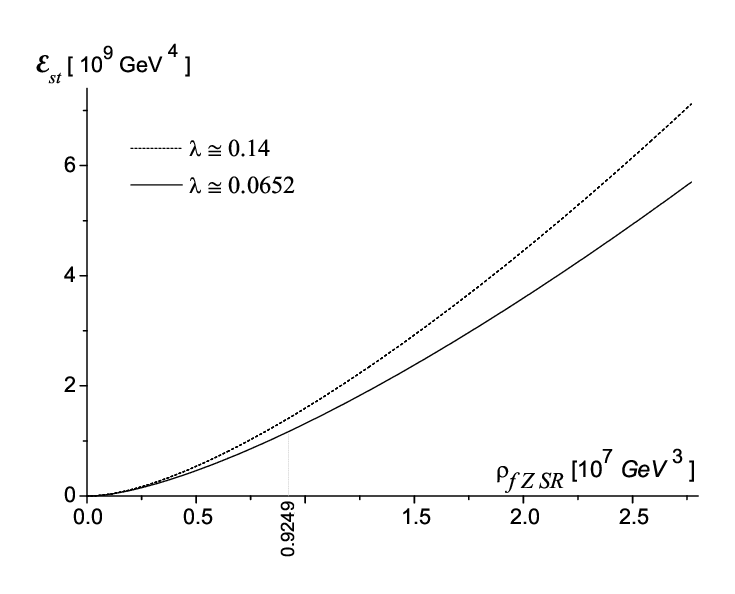}
\end{center}
\caption{ {\it The energy density of the Wbgfms configuration ${\cal E}_{st}(\varrho_{f Z \; SR}) = {\cal U}^{ef}_{f}(\vartheta=0,\;
\delta \neq 0, \; \varrho_{f Q \; SR} = 0)$ (see Eq.(\ref{Pdz_82})).
}
}
\end{figure}

From Eqs.(\ref{Pdz_78}) and (\ref{Pdz_73}), it is clear that the
appearance of $\varrho_{f Z \; SR} > 0$ (so
$\zeta > 0$) leads to the instability in the
$W^{\pm}_{\mu}$ sector only if
\begin{eqnarray}
\label{Pdz_83} \zeta^{3} (\varrho_{f Z \; SR}) > \frac{\sqrt{g^{2} + g'^{2}} \, \varrho_{f Z \; SR}}{g^{2}} \; ,
\end{eqnarray}
which is connected with the fact that then $m_{\tilde{W}^{\pm}}^{2} < 0$ \cite{Dziekuje_Jacek_nova_2}.
When the equality $\zeta^{3} (\varrho_{f Z \; SR}) = \sqrt{g^{2} + g'^{2}} \, \varrho_{f Z \; SR}/g^{2}$ is taken into account, we obtain the relationship
between $\lambda_{max}$ and $\varrho_{f Z \, SR \, max}$, where
$\lambda_{max}$ is the value of $\lambda$ and $\varrho_{f Z \, SR
\, max}$ is the value of $\varrho_{f Z \; SR}$ for which we have
$m_{\tilde{W}^{\pm}}^{2} = 0$. The region of stable Wbgfms
configurations with $\zeta \neq 0$ is on and below the
$\varrho_{f Z \, SR  \, max}(\lambda_{max})$ boundary curve  (see Figure~7a).
\\
For the weak charge density fluctuation $\varrho_{f Z \; SR} < \varrho_{f Z \; SR}^{limit} \equiv ( g^2 + g'^{\,2})\,  v^{3}/(8\,g)$
$\approx
%
%
1.585 \times 10^6  \; $ ${\rm GeV}^{3}$,
this configuration of fields is stable for an arbitrary $\lambda$ (see Figure~7a).
For values of $\varrho_{f Z \; SR}$ bigger than $\varrho_{f Z \; SR}^{limit}$,
%
%
the Wbgfms configuration is unstable for a given $\lambda$ above
a certain value of $\varrho_{f Z \; SR}$,
which is equal to
\begin{eqnarray}
\label{rhomax} \varrho_{f Z \; SR max} = \frac{8 g^2
\lambda^{\frac{3}{2}}  \,(g^{2} + g'^{\,2})\, v^{3}}{\left[ 16
g^{2} \lambda - (g^{2} + g'^{\,2})^{2} \right]^{\frac{3}{2}}} \; .
\end{eqnarray}
For $\lambda < \lambda_{limit} \equiv (g^{2} + g'^{\,2})^{2}/(16 g^{2}) \approx
%
%
0.0451$,  the Wbgfms configuration is stable for all values of
$\varrho_{f Z \; SR}$ (see Figure~7a).

\subsection{The mass of the Wbgfms configuration}

\label{mass of neutral bgfms}

Let us examine the mass of the droplet of the Wbgfms configuration induced by the non-zero weak charge density fluctuation $\varrho_{f Z \; SR}$
\begin{eqnarray}
\label{M3 config}
\!\!\!\! M_{i^{3}_{f}} = \frac{4}{3} \, \pi \, r_{i^{3}_{f}}^{3} \, {\cal E}_{st}(\varrho_{f Z \; SR}) \;  \; {\rm and} \;\;  M_{i^{3}_{f}} =  \pm \, i^{3}_{f} \, M_{i^{3}_{f} = 1}  \, ,
\end{eqnarray}
where ${\cal E}_{st}(\varrho_{f Z \; SR})$ is given by Eq.(\ref{Pdz_82}) and the sign ``+'' is for $i^{3}_{f} > 0$ and  ``-'' for $i^{3}_{f} < 0$. Because of the Pauli exclusion principle used for the fermionic fluctuations, only $i^{3}_{f}=\pm 1/2$ or $\pm 1$   (see Table) inside one droplet  are possible  (except in cases where the consecutive fermionic fluctuations occupy their higher energy states). Here, $r_{i^{3}_{f}}$ is the ``radius of the weak charge density fluctuation'' determined by the
%
%
weak isotopic charge fluctuation inside the Wbgfms configuration in the thin wall approximation
\begin{eqnarray}
\label{ri3 radius of Mi3}
i^{3}_{f} = \frac{4}{3} \, \pi \, r_{i^{3}_{f}}^{3} \, \varrho_{f Z \; SR} \; .
\end{eqnarray}
The radius $r_{i^{3}_{f}}$ is the function of $i^{3}_{f}$.
The value of $|i^{3}_{f}|$ inside one droplet can possibly be more than 1 for the composite fermion fluctuation only \cite{composite_fermion}.
%
%
\begin{figure}[here]
\begin{center}
\includegraphics[angle=0,width=75mm]{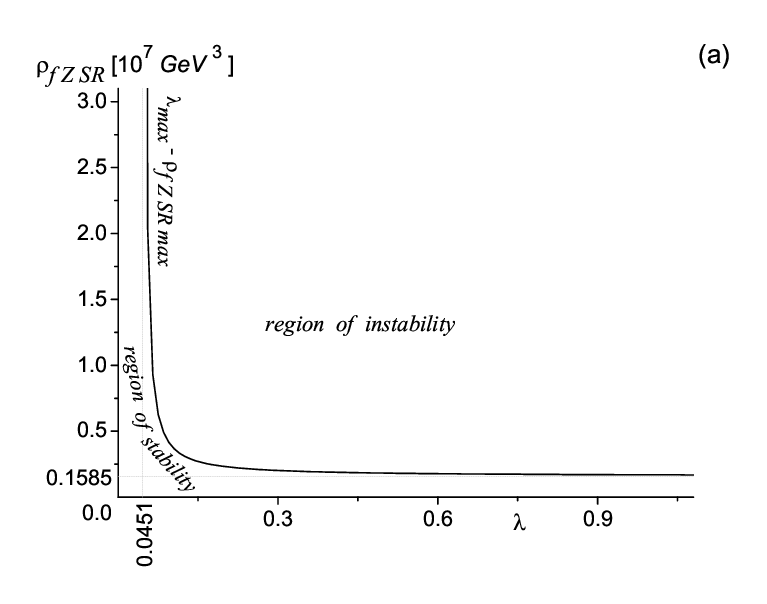}
\includegraphics[angle=0,width=75mm]{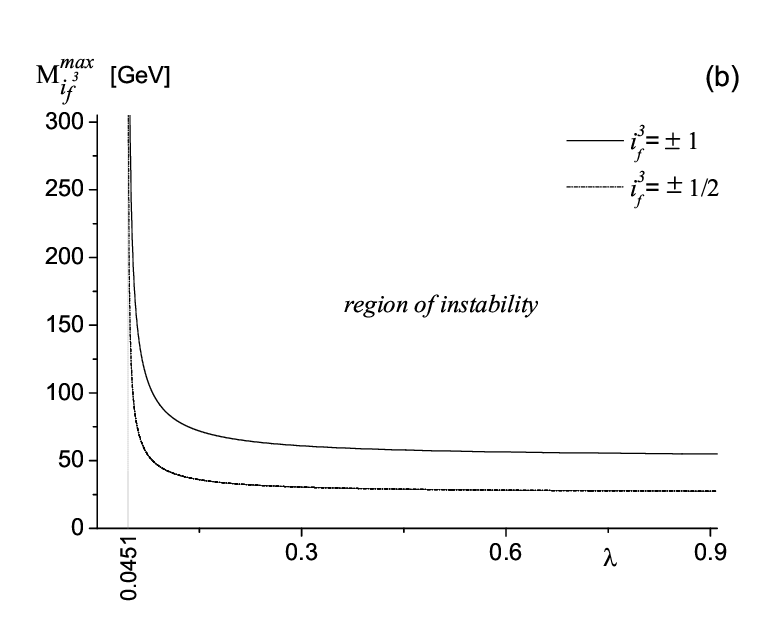}
\caption{ {\it
Panel: (a) The partition of the
($\lambda,\varrho_{f Z \; SR}$) plane into the regions of
stability and instability of the Wbgfms configurations
with $\varrho_{f Z \; SR} \neq 0$.
The region of stable
Wbgfms configurations lies on and below
$\varrho_{f Z \, SR max}(\lambda_{max})$ boundary curve, where
$\lambda_{max}$ is the value of $\lambda$ and $\varrho_{f Z \, SR
max}$ is the value of $\varrho_{f Z \; SR}$ for which
$m_{W^{\pm}}^{2} = 0$. The limiting values $\varrho_{f Z \; SR}^{limit}  \approx
0.1585 \times 10^7  \; $ ${\rm GeV}^{3}$ and $\lambda_{limit} \approx
0.0451$ are shown.
\newline
(b) The upper mass $M^{max}_{i^{3}_{f}}$ (according to the stability of the Wbgfms configuration
in the ${W}^{\pm}$ sector) with $\varrho_{f Z \; SR} \neq 0$
as a function of $\lambda=\lambda_{max}$,
where $m_{\tilde{W}^{\pm}}^{2} = 0$ for
points ($\lambda_{max}$, $M^{max}_{i^{3}_{f}}$) which lie on the curve.
The region of possible Wbgfms configurations is on and below the $M^{max}_{i^{3}_{f}}(\lambda_{max})$ boundary curve. Two such curves, the first one for $i^{3}_{f} = \pm 1$
and the second one for $i^{3}_{f} = \pm 1/2$
are plotted.
For $\lambda \rightarrow \infty$
$M^{max}_{i^{3}_{f} = \pm 1} \approx 52.277$ GeV
and $M^{max}_{i^{3}_{f} = \pm 1/2} \approx 26.138$ GeV,
respectively.
}}
\end{center}
\end{figure}

According to the stability of the Wbgfms configuration in respect of the $W^{\pm}$ sector, we can also obtain the upper limit
$M^{max}_{i^{3}_{f}}$ for the value of the mass $M_{i^{3}_{f}}$.
The region of the stability of possible Wbgfms configurations lies
on and below the proper $M^{max}_{i^{3}_{f}}(\lambda_{max})$ boundary curve (see Figure~7b). Two such curves are presented,
one for the function  $M^{max}_{i^{3}_{f}= \pm 1}(\lambda)$ and the other for $M^{max}_{i^{3}_{f} = \pm 1/2}(\lambda)$.
In principle, the value of $\lambda$ can be readout from the particular curve when the experimental value
%
%
of the mass $M^{max}_{i^{3}_{f}}$ is known. \\

{\bf Note}: It is not difficult to see that
$\varrho_{f Z \; SR} \longrightarrow 0$ (which implies $\zeta
\longrightarrow 0 $ and $\delta \longrightarrow v$) entails ${\cal E}_{st}(\varrho_{f Z \; SR}) \longrightarrow 0$ for the energy density (\ref{Pdz_82})  of the limiting  Wbgfms configuration.
The double limit $\varrho_{f Z \; SR} \longrightarrow 0$ and $i^{3}_{f} \longrightarrow 0$  is the only possibility for obtaining the weakly uncharged Wbgfms configuration.
%
%
From Figure~7b it can be noticed that for the established value of  $\lambda > \lambda_{limit}  \approx 0.0451$ and with $i^{3}_{f} \longrightarrow 0$, the maximal mass $M_{i^{3}_{f}}$ of the Wbgfms configuration, which lies on the boundary curve $M^{max}_{i^{3}_{f}}(\lambda_{max})$,  also tends to zero. Thus, in this case in the double limit $\varrho_{f Z \; SR} \longrightarrow 0$ and $i^{3}_{f} \longrightarrow 0$, the Wbgfms configuration becomes  necessarily  massless for $\lambda > \lambda_{limit}$ (for $\lambda \leq \lambda_{limit}$ this would be not necessarily the case). \\
At the same time, from Figure~5a-b we notice that for $\varrho_{f Z \; SR} \longrightarrow 0$,
%
%
the Wbgfms configuration reproduces some characteristics of the uncharged $\varrho_{f Z \; SR} = 0$
configuration, e.g. the masses of the (composite) boson fields and the lack of self-consistent gauge fields. Nevertheless, even for an
infinitesimally small value of $\varrho_{f Z \; SR}$, the value of the self-consistent field $\delta$ is different from zero and tends in the limit to $v$.
%
%
Thus, for $\lambda > \lambda_{limit}$ (which will be suggested later on) and for
$\varrho_{f Z \; SR} \longrightarrow 0$, $i^{3}_{f} \longrightarrow 0$,  the particles interacting with this massless Wbgfms configuration can perceive the fields that are inside a Wbgfms droplet with their SM values of couplings.

\section{The intersections of EWbgfms and Wbgfms configurations}

\label{intersection}

Let us start with the electrically charged EWbgfms configuration
%
%
with a matter electric charge fluctuation equal to $q_{f} = 2$ (analysis for $q_{f} = -2$ would be the same) and a minimal mass $M^{min}_{q_{f} = 2}$.
Now, let us pose the question on the configuration of the nearest Wbgfms droplet with $\varrho_{f Z \; SR} \neq 0$
%
%
that arises after the decay of this minimal mass EWbgfms configuration with $\varrho_{f Q} \neq 0$.
%
%
The solution with a particular value of $\lambda$ can be found as the  point of the  intersection of the function of the minimal masses $M^{min}_{q_{f}}(\lambda)$ of EWbgfms configurations (presented on Figure~4b) with the function of the maximal masses $M^{max}_{i^{3}_{f}}(\lambda)$
of Wbgfms configurations (presented on Figure~7b).
Six such solutions can be seen in Figure~8a.
%
%
\begin{figure}[top]
\begin{center}
\includegraphics[angle=0,width=71mm]{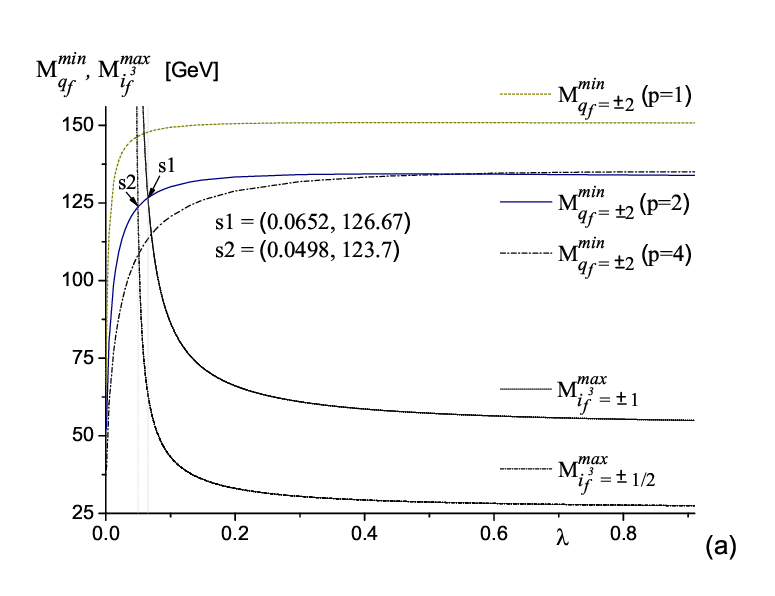}
\includegraphics[angle=0,width=71mm]{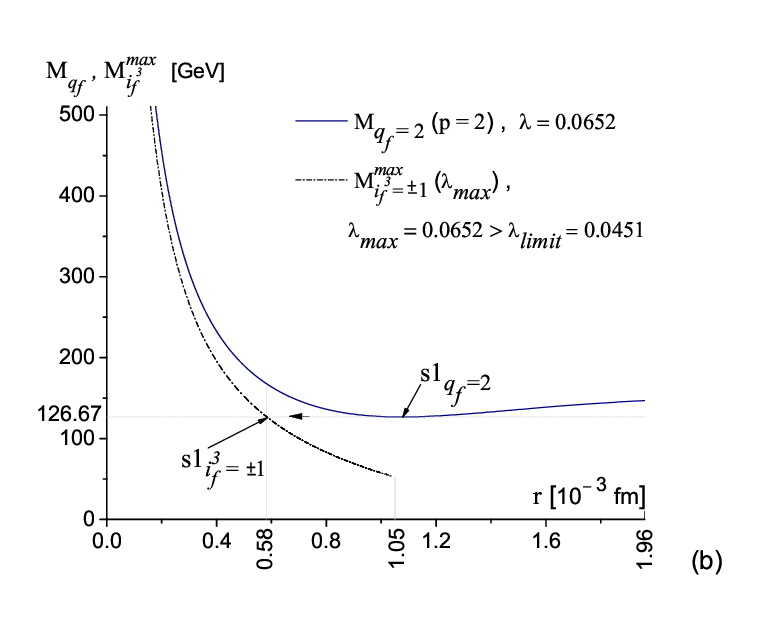}
\caption{ {\it
Panel (a): The intersections of the curves of the minimal masses $M^{min}_{q_{f}}(\lambda)$ of the EWbgfms configurations
(presented on Figure~4b) with the curves of the maximal masses $M^{max}_{i^{3}_{f}}(\lambda)$ of the
Wbgfms configurations (presented on Figure~7b).
\newline
(b): The EWbgfms configurations
with the mass $M_{q_{f}=2}$ as a function of the radius $r_{q_{f}=2}$ and the Wbgfms configurations with the upper
mass $M^{max}_{i^{3}_{f}= \pm 1}(r_{i^{3}_{f}})$ (according to the stability in the ${W}^{\pm}$ sector) as a function of the radius
$r_{i^{3}_{f} = \pm 1}$.
The ``decomposition'' of the particular solution $s1$ found in Figure~8a is shown. Two points, i.e. $s1_{q_{f}=2}$ with $M^{min}_{q_{f}=2}(r_{q_{f}}) \approx 126.67$ GeV on the curve $M_{q_{f}=2}(r_{q_{f}})$ and $s1_{i^{3}_{f}=\pm 1}$ with the same mass on the curve $M^{max}_{i^{3}_{f}=\pm 1}(r_{i^{3}_{f}})$ correspond to one point $s1$ in Figure~8a.
The right cut $r^{int}_{q_{f}=2} \sim 1/m_{Z} \approx 0.00196
\; fm$ on the curve $M_{q_{f}=2}(r_{q_{f}})$ is connected with the thin wall approximation, whereas for the curve
$M^{max}_{i^{3}_{f}=\pm 1}(r_{i^{3}_{f}})$ the maximal value of $r_{i^{3}_{f}= \pm 1} \approx 0.00105 \; fm$  follows from the fact that for $\lambda \rightarrow \infty$ the limiting, lowest possible value of $\varrho_{f Z \; SR}$ for these upper mass configurations is equal to $\varrho_{f Z \; SR}^{limit} \approx 1.585 \times 10^6  \; $ $GeV^{3}$ (see Figure~7a).
}}
\end{center}
\end{figure}\\
%
%
%
%
%
%
%
The estimates obtained for the mass of the observed neutral state in the LHC experiment \cite{cms_1,cms_2}, \cite{atlas} are in case of the CMS detector equal currently to
$126.2$ $\pm 0.6$ (stat) $\pm 0.2$ (syst) GeV for its $ZZ^{(*)} \rightarrow 4 \ell$ ($\ell= e$ or $\mu$) decay channel \cite{cms2}
and in case of the ATLAS detector equal to $126.6$ $\pm 0.3$ (stat) $\pm 0.7$ (sys) GeV in the $\gamma \gamma$ channel or
$123.5$ $\pm 0.8$ (stat) $\pm 0.3$ (sys) GeV in the $ZZ^{(*)} \rightarrow 4 \ell$ channel  \cite{atlas2}.
%
%
%
%
Therefore, from the estimates obtained in the LHC experiment, only two solutions for the intersection of functions $M^{max}_{i^{3}_{f}}(\lambda)$ (one for $i^{3}_{f}=\pm 1/2$ and the other for $\pm 1$) with the function of the minimal masses  $M^{min}_{q_{f}=2}(\lambda)$ for $p=2$ remain. These are the solutions $s1$ and $s2$,  which are discussed below.
\\
%
%

For the solution $s1$ in Figure~8a, we obtain  $\lambda \approx 0.065187 \approx 0.0652$
and $M^{min}_{q_{f}= \pm 2} = M^{max}_{i^{3}_{f}=\pm 1} \approx 126.67$ ${\rm GeV}$. Firstly, let us write down the characteristics of the electrically charged EWbgfms configuration
with $\varrho_{f Q} \neq 0$ (see Eq.(\ref{Pdz_58}) and Figure~1a).
Thus, the electric charge density fluctuation is equal to
%
%
$\varrho_{f Q} =  2.965  \times 10^6$ ${\rm GeV}^3$ (compare Figure~2a) and the energy density  (Figure~3b)
is equal to
%
%
${\cal E}_{st}(\varrho_{f Q}) \approx 1.878 \times 10^8$
${\rm GeV}^4$.
For $q_{f}= 2$ the radius of the electrically charged EWbgfms configuration is equal to $r_{q_{f}} \approx  0.00107 \; fm$ (see Figures~4a and 8b).
For  $\varrho_{f Q} =  2.965  \times 10^6$ ${\rm GeV}^3$ the mass $m_{\tilde{Z}} \approx 124.128$ ${\rm GeV}$ inside the droplet of the EWbgfms configuration
is the biggest one (see Figure~3a);
%
%
hence, the interaction range $r^{int}_{q_{f}}$ inside the droplet is of the order $r^{int}_{q_{f}} \sim 1/m_{\tilde{Z}} \approx 0.00159  \; fm$ and because the ratio $r_{q_{f}}/r^{int}_{q_{f}} \approx 0.675 <1$, it is reasonable to use the thin wall approximation. \\
The other, i.e. the electrically neutral Wbgfms configuration of the solution $s1$ with the non-zero weak charge density fluctuation
$\varrho_{f Z \; SR max} \approx 9.249 \times 10^6$ ${\rm GeV}^3$,
has the energy density
${\cal E}_{st}(\varrho_{f Z \; SR max}) \approx 1.172 \times 10^9$ ${\rm GeV}^{4}$ (Figure~6). For $i^{3}_{f}=\pm 1$ its radius
is equal to $r_{i^{3}_{f}} \approx  0.000583 \; fm$ (see Figure~8b).
For this value of $\varrho_{f Z \; SR max}$,
the mass $ m_{\tilde{Z}}= 165.064 $ ${\rm GeV}$
(see Figure~5b) is the biggest one ($m_{\tilde{\varphi}_{f}} = 160.071$ ${\rm GeV}$);
hence, the interaction range $r^{int}_{i^{3}_{f}}$ inside the droplet of the Wbgfms configuration is of the order $r^{int}_{i^{3}_{f}} \sim 1/m_{\tilde{Z}} \approx 0.0012
\; fm$. Thus, because the ratio $r_{i^{3}_{f}=1}/r^{int}_{i^{3}_{f}} \approx 0.488 <1$,  it is reasonable to use the thin wall approximation.\\

The transition from the electrically charged EWbgfms configuration (state $s1_{q_{f}=2}$) to the uncharged Wbgfms configuration (state $s1_{i^{3}_{f} = \mp 1}$) is presented in Figure~8b.
These two points are represented by one solution $s1$ on the $\lambda - M$ plane in Figure~8a. We interpret the electrically uncharged Wbgfms configuration represented by the point $s1_{i^{3}_{f} = \mp 1}$ as the candidate for the neutral state of the mass $ \sim 126.5$  GeV recently observed  in the LHC experiment.  \\
{\bf Remark}: In the presented paper, the masses of the states $s1_{q_{f}=+2}$ and $s1_{i^{3}_{f}}$ (or $s2_{q_{f}=+2}$ and $s2_{i^{3}_{f}}$) are equal.
Yet, the mass splitting between the states $s1_{q_{f}=+2}$ and $s1_{i^{3}_{f}}$
(or $s2_{q_{f}=+2}$ and $s2_{i^{3}_{f}}$) could be of the 10 MeV order, which is in agreement with the value of the decay width of the $126.5$ GeV boson state observed in the LHC experiment  \cite{Barger-Ishida-Keung,CMS-do-Barger-Ishida-Keung} (also Section~\ref{process}).
Then, examples a11-b2 which follow, which have on their right hand sides the dielectron events plus neutrinos, are from this point of view not excluded by the present LHC experiment. \\
\\
The examples of the processes connected with $s1$
%
%
are as follows.
\\
\\
For $s1_{q_{f}=+2}$ and $s1_{i^{3}_{f}=\pm 1}$, which are the leptonic states: \\
{\bf a11}) $\; p+p \rightarrow (s1_{q_{f}=+2}) + X + 2 \nu\;
$ and then
$\; (s1_{q_{f}=+2}) \rightarrow (s1_{i^{3}_{f} = - 1})
+ 2 \nu + 2 e^{+} $\\
{\bf a12}) $\; p+p \rightarrow (s1_{q_{f}=+2}) + X + 2 \nu\;
$ and then
$\; (s1_{q_{f}=+2}) \rightarrow (s1_{i^{3}_{f} = + 1})
+ 2 {\bar \nu} + 2 e^{+} $\\
\\
For $s1_{q_{f}=+2}$ and $s1_{i^{3}_{f}=\pm 1}$, which are the barionic states:\\
{\bf a2}) $\; p+p \rightarrow (s1_{q_{f}=+2}) + X +  ({}^{\nu \ell^{+}}_{{\bar \nu}\ell^{-}} )\;
$ and then
$\; (s1_{q_{f}=+2}) \rightarrow (s1_{i^{3}_{f} = \mp 1})
+ 2 \nu + 2 e^{+} $.\\
\\
Here $\ell$ is the electron or muon
and $X$ signifies some jets.\\

For the solution $s2$ in Figure~8a, we obtain correspondingly
$\lambda \approx 0.04977 \approx 0.0498$
%
%
and $M^{min}_{q_{f}= \pm 2} = M^{max}_{i^{3}_{f}=\pm 1/2} \approx 123.7$ ${\rm GeV}$.
%
%
The characteristics of the electrically charged EWbgfms configuration are as follows:
%
%
$\varrho_{f Q} \approx  2.615  \times 10^6$ ${\rm GeV}^3$,
%
%
${\cal E}_{st}(\varrho_{f Q}) \approx 1.618 \times 10^8$
${\rm GeV}^4$ and for $q_{f}= 2$ the radius
of the droplet is equal to $r_{q_{f}} \approx  0.00112 \; fm$.
For this value of $\varrho_{f Q}$
%
%
the mass $m_{\tilde{Z}} \approx 121.940$ ${\rm GeV}$ is the biggest one;
hence, the interaction range $r^{int}_{q_{f}}$ inside the droplet of the EWbgfms configuration is of the order $r^{int}_{q_{f}} \sim 1/m_{\tilde{Z}} \approx  0.00162 \; fm$. Because $r_{q_{f}}/r^{int}_{q_{f}} \approx 0.692 <1$, it is reasonable to use the thin wall approximation.
The characteristics of the electrically neutral Wbgfms configuration are as follows:
%
%
$\varrho_{f Z \; SR max} \approx 5.477 \times 10^7$ ${\rm GeV}^3$
with
%
%
${\cal E}_{st}(\varrho_{f Z \; SR max}) \approx 1.355$
$ \times 10^{10}$
${\rm GeV}^{4}$ and for $i^{3}_{f}= \pm 1/2$ we obtain $r_{i^{3}_{f}} \approx  0.000256 \; fm$.
For this value of $\varrho_{f Z \; SR max}$,
the mass $m_{\tilde{Z}}=298.621$
${\rm GeV}$ is the biggest one ($m_{\tilde{\varphi}_{f}} \approx 253.036$ ${\rm GeV}$);
hence, $r^{int}_{i^{3}_{f}} \sim 1/m_{\tilde{Z}} \approx 0.000661  \; fm$. Because $r_{i^{3}_{f}}/r^{int}_{i^{3}_{f}} \approx 0.387 <1$,  it is reasonable to use the thin wall approximation.
The exemplary processes for the $s2$ case (see Figure~8a) are as
follows.
\\
\\
For $s2_{q_{f}=+2}$ and $s2_{i^{3}_{f}=\pm 1/2}$, which are the leptonic states:\\
{\bf b11}) $\; p+p \rightarrow (s2_{q_{f}=+2}) + X + 2 \nu\;
$ and then
$ \; (s2_{q_{f}=+2}) \rightarrow (s2_{i^{3}_{f} = - 1/2})
+ \nu + 2 e^{+} $ \\
{\bf b12}) $\; p+p \rightarrow (s2_{q_{f}=+2}) + X + 2 \nu\;
$ and then
$ \; (s2_{q_{f}=+2}) \rightarrow (s2_{i^{3}_{f} = + 1/2})
+ {\bar \nu} + 2 e^{+} $ \\
\\
For $s2_{q_{f}=+2}$ and $s2_{i^{3}_{f}=\pm 1/2}$, which are the barionic states:\\
{\bf b2}) $\; p+p \rightarrow (s2_{q_{f}=+2}) + X + ({}^{\nu \ell^{+}}_{{\bar \nu}\ell^{-}} )
$ and then
$ \, (s2_{q_{f}=+2}) \rightarrow (s2_{i^{3}_{f} = \mp 1/2})
+ 2 \nu +2 e^{+} $.\\
\\
Some of the above processes look like the lepton number violation  (i.e. a12, b11 and b12), but if $s_{q_{f}}$ and $s_{i^{3}_{f}}$ are  leptonic states, they are not really of this type. \\
If the LHC state can be either a barionic or leptonic one, then the  $q_{f}=2$  possibility is chosen  only on the basis of the observed mass. Next, if the droplets of the bgfms configurations are leptonic,  then the states with $|q_{f}| > 2$ are (by the Pauli exclusion principle) possible only if the consecutive fermionic fluctuations are in higher energy states. Nevertheless, in both cases, the barionic and leptonic, the particular function $M^{min}_{q_{f}}(\lambda)$
for the EWbgfms configurations with $|q_{f}| > 2$ intersects with the functions $M^{max}_{i^{3}_{f}}(\lambda)$ of the Wbgfms configurations for higher masses and these solutions have not yet been observed in
an LHC experiment.  \\
%
%
Let us consider the case when
the bgfms configurations $s_{q_{f}}$ and $s_{i^{3}_{f}}$ are occupied by two (electrically charged and uncharged, respectively) fermionic fluctuations with opposite spin projections. In addition to the scalar fluctuation $\varphi_{f}$, there are four gauge self-fields inside the
%
%
configuration given by Eq.(\ref{Pdz_58}) and three inside the
%
%
configuration given by Eq.(\ref{Pdz_77})).
Thus, for the particular configuration of the ground fields given by Eq.(\ref{Pdz_58}), its EWbgfms $s_{q_{f}}$ droplet can have spin zero (and zero to four for its excitations). Meanwhile, for the particular configuration of the ground fields given by Eq.(\ref{Pdz_77}), its Wbgfms $s_{i^{3}_{f}}$ droplet can have spin zero (and zero to three for its excitations  \cite{helicity_of_W_Bilenkii,helicity_of_W_Gounaris,helicity_of_W_Fleischer,helicity_of_W_Bella}).
Indeed, because $s_{i^{3}_{f}=\pm 1}$ is the ground state configuration,  hence the self-consistent field $Z_{0}$ exists only inside its droplet  (see Eq.(\ref{Pdz_77})),
which
belongs to the spin zero subspace of the 3-dimensional rotation group.
Thus, the $s_{i^{3}_{f}=\pm 1}$  ground configuration of fields, which consists of two opposite spin fermionic fluctuations, the scalar fluctuation $\varphi_{f}=\delta$ and spin zero $Z_{0}=\zeta$, has a  spin equal to zero. (When boosted the $Z$ self-field is longitudinally
polarized, i.e. its spin is equal to one with a spin projection equal to zero.)
%
%
Next, from the point of view of the possible value of the spin of the Wbgfms configuration,
considerations similar to the  ones above (for two fermionic fluctuations) lead to the conclusion that states $s_{i^{3}_{f}=\mp 1/2}$ in b11, b12 and b2 with quantum numbers for fermionic fluctuation like those in the Table {\it are excluded} by the LHC experiment as they consist of one fermionic fluctuation only  thus having a half spin value.
\\
We see that only cases a11, a12 and a2 are possible
and thus the present day experiments have selected
the state  $s1_{i^{3}_{f} = \mp 1}$ with mass $M^{max}_{i^{3}_{f}=\mp 1} \approx 126.67$ ${\rm GeV}$ for $\lambda \approx 0.0652$ and
rejected the state $s2_{i^{3}_{f} = \mp 1/2}$ with mass $M^{max}_{i^{3}_{f}=\mp 1/2} \approx 123.7$ ${\rm GeV}$ for $\lambda \approx 0.0498$.
However, the basic fields that induce the bgfms configurations of fields are (in this model) the fermionic fluctuations; hence, one could think that the states $s_{q_{f}=2}$ and $s_{i^{3}_{f}=\mp 1}$ are  leptonic states (think of some models of a neutron in which the
neutron is a composition of a barionic proton and a fermionic electron
\cite{Santilli_1,Santilli_2}). In this case, only the possibility of a leptonic state $s1_{i^{3}_{f} = \mp 1}$ remains, which is exemplified by  processes a11 and a12. Otherwise, the barionic states exemplified by processes a2 remain with the configuration $s1_{i^{3}_{f}=\mp 1}$ suggested as the solution for the state observed in the LHC experiment.\\
In Figure 8a three pairs of neighbouring solutions can be noticed.
Nevertheless, whether besides the experimentally noticeable state $s1_{i^{3}_{f}=\pm 1}$, the neighbouring solution $s2_{i^{3}_{f}=\mp 1/2}$ together with the remaining ones have been also observed \cite{cms3_1,cms3_2,cms3_3,cms3_4}
as more shallow resonances and not as the statistical flukes in the data only,
remains an open question.
The reason is that
in such a case $\lambda$ gains two additional indexes, i.e. $\lambda \rightarrow \lambda_{p,i^{3}_{f}}$, where the {\it electric charge to hypercharge ratio index} $p$, Eq.(\ref{Pdz_57}), numbers the EWbgfms configurations and the {\it weak isotopic charge} $i^{3}_{f} = \mp \frac{1}{2}, \mp 1$ numbers the Wbgfms ones.\\
Thus, in Figure 8a for
each $p$, where
$p=1, 2$ and 4, one pair ${ s_{i^{3}_{f}=\mp 1/2} \choose \!\!\! s_{i^{3}_{f} = \mp 1} }$ of the neighbouring solutions:
\begin{eqnarray}
\label{pairs of states}
& & \!\!\!\!\!\!\!\! \!\!\!\!\!\!\!\! \!\!\!\!\!\!\!\! \!\!\!\!\!\!\!\!
{ (0.0512, 108.79) \choose (0.0705, 114.91) }  \, , \; \nonumber \\
{ (0.0498, 123.7) \choose  (0.0652, 126.67) } \;\;  & {\rm and} & \;\;  { (0.0484, 146.33) \choose  (0.0593, 147.4) }\, ,
\end{eqnarray}
respectively, can be noticed, where for each of the six solutions the values of $\lambda$ and $M^{max}_{i^{3}_{f}}$ [${\rm GeV}$] are given.  \\
The central column in Eq.(\ref{pairs of states}) is ${ s2_{i^{3}_{f}=\mp 1/2} \choose \!\!\! s1_{i^{3}_{f} = \mp 1} }$.
It is easy to notice that the algebraic mean of the mass of two central neighbouring solutions $s1_{i^{3}_{f}=\pm 1}$ and $s2_{i^{3}_{f}=\mp 1/2}$ is equal to 125.185 GeV. This value is consistent with the mean mass of the configurations observed in the first run of the LHC experiment (with higher than $5 \sigma$ significance of the observed
excess over the expected background \cite{PDG-2014-2015}).
Yet, it has to be also noticed that the values
in the third column in Eq.(\ref{pairs of states}) lie in the vicinity of the
events recorded in the CMS experiment at a mass of approximately 145 GeV
with a statistical significance of
$\sim 3 \sigma$ above background expectations
\cite{cms3_1,cms3_3,Khallil-Moretti}.
\\

Finally, it is not difficult numerically to check that for all
Wbgfms configurations that lie on their boundary curve $M^{max}_{i^{3}_{f}}(\lambda_{max})$
and have a particular value of the weak isotopic charge fluctuation $i^{3}_{f}$,
%
%
the relation
\begin{eqnarray}
\label{delta from i3}
\frac{4}{3} \, \pi \, r_{i^{3}_{f}}^{3} \, \delta^{3}/(3 \, \pi) \approx |i^{3}_{f}| \;
%
%
\end{eqnarray}
is fulfilled (up to the fourth digit after the decimal point).
The mass of the droplet calculated with $\delta$ obtained from the perfect equality in Eq.(\ref{delta from i3}) with the (non self-consistent) use of Eq.(\ref{E min rhoQ}) agrees with
$M^{max}_{i^{3}_{f}}$ up to the seventh digit after the decimal point.
%
%
Thus, the relation (\ref{delta from i3}) is also fulfilled by the configurations $s1_{i^{3}_{f} = \mp 1}$ (and  e.g. $s2_{i^{3}_{f} = \mp 1/2}$ also).
%
%
The Wbgfms configuration $s1_{i^{3}_{f} = \mp 1}$ is the successor of the EWbgfms configuration $s1_{q_{f}=2}$.  In Figure~8a these configurations overlap. Both  have a mass equal to
%
%
$126.67$ GeV, which (besides the spin zero) has been interpreted above as the signature of the LHC state. In this way both the charge $q_{f}=2$ and $i^{3}_{f}=\pm 1$ are discreetly chosen.
Thus,  it is
suggested by Eq.(\ref{delta from i3}) that in the one parametric $\varrho_{f Z \; SR} \neq 0$ case (see
Eqs.(\ref{Pdz_73})-(\ref{Pdz_72})), the quantization
%
%
$i^{3}_{f} = \pm 1$
is an artefact of the self-consistency conditions given by Eqs.(\ref{Pdz_69})-(\ref{Pdz_70}).
The analysis of condition (\ref{delta from i3}) will be discussed in a following paper.
%
%
%
%
%

\subsection{The decay of the $s1_{i^{3}_{f} = \mp 1}$ droplet}

\label{process}

In the full-self consistent field theory, fields have the same type of couplings as their counterparts in the perturbative quantum field theory. This is the case of e.g. the self-consistent electrodynamics and one of its outcomes is the derivation of the Lamb shift by Barut and Kraus
\cite{bib_B-K-1}.
Although the presented CGSW model treats the self-consistent field and the wave self-field of excited states differently, a self-field is in reality one object (on the  ground state, i.e. in the droplet of a bgfms configuration,  only self consistent fields are present). Thus, both the self-consistent field and the wave self-field in CGSW have the same type of couplings as their counterparts in the GSW model.\\
The self-consistent electrically uncharged Wbgfms configuration $s1_{i^{3}_{f} = \mp 1}$ is
the resonance via the weak interactions only
and can disintegrate through the simultaneous decay or radiation of its constituents.
%
%
%
In a droplet of a Wbgfms configuration of fields induced by $\varrho_{f Z \; SR} \neq 0$ (with $\varrho_{f Q \; SR} = 0$),
the self-consistent fields $\varphi_{f}$ and $Z$ (see Eq.(\ref{Pdz_77}))
are present in addition  to the background fermionic  fluctuations.
%
%
Then, only $\delta$ of $\varphi_{f}$ and the time component $\zeta$ of $Z$ are different from zero.
%
%
(Due to $\varrho_{f Q \; SR} = 0$ and $m_{A}=0$, the electromagnetic self-field $A$ is totally absent even in the excitation; however, the pair $W^{+}-W^{-}$ of the self-fields can appear in the excitation.)
The self-consistent fields
%
%
are the initial ones that take part in the decay of the Wbgfms configuration.
For each initial self-consistent field the calculation of the coherent transition probability is performed separately (i.e. for $\varphi_{f}=\delta$ and $Z_{0}$) and then the decay of the droplet of the Wbgfms configurations is calculated in accordance with the following scenario. Firstly, there appears the decay of the coupling of the self-consistent field $\zeta$ of $Z_{\mu}$ to the basic fermionic field followed by the decay of $\zeta$ (which is very rapid in SM). Then, (for $\varrho_{f Z \; SR} \longrightarrow 0$, $i^{3}_{f} \longrightarrow 0$ and $\lambda > \lambda_{limit}$) the fields configuration of the droplet decays (see Note in Section~\ref{mass of neutral bgfms}). In this limit, the particles interacting with the configuration can perceive, with the SM values of couplings, fields that are inside the droplet. This leads to the decay of the self-consistent field $\delta$ of the scalar fluctuation  $\varphi_{f}$ with the decay rate of the same order as predicted for the SM Higgs particle. Thus, roughly speaking, the decay width of the Wbgfms configuration shall be of the order of a few MeV.
Finally, only longer-lived particles are detected in the detector.


%
%

\subsection{Transparency of the uncharged bgfms configuration to  electromagnetic radiation}

\label{photon transparency}

In Section~\ref{neutral} it was noted that the effective mass $m_{A}$ of the electromagnetic self-field $A$ inside the droplet of an  electrically uncharged Wbgfms configuration
%
%
is equal to zero. Although the electromagnetic self-field is totally absent in this bgfms configuration (see Section~\ref{neutral}), zeroing of the effective mass and $\varrho_{f Q} = 0$ are important for the photons that are external ones  (see Introduction). The reason is that the formal form of the equation of motions (\ref{Pdz_14})-(\ref{Pdz_16}) is also  true for the external gauge fields penetrating the discussed bgfms configuration. Thus, the Wbgfms configuration is transparent for the external electromagnetic radiation. \\
Now, let us suppose that the matter is extremely dense, as could happen in the mergers of neutron stars. Then the difference between the inward structure of the nucleon and the inward structure of the droplet of the Wbgfms configuration may be a supporting impulse to initiate the relativistic shock. That is, the abrupt transition of the neutron matter during the collapse of star mergers could cause the transition to matter of Wbgfms droplets, which are transparent to the gamma radiation that is produced within the gamma-ray bursts (GRB) explosion. This can lead to the appearance of an alternative source of energy that can help the gamma-ray burst \cite{MarekJacek}. This would also be the reason for the recently observed lack of correlations between gamma-ray bursts and the neutrino fluxes (present in the standard model \cite{dziekuje_za_DLS2} and) directed from them
\cite{bursts_lack_neutrino_LoSecco,bursts_lack_neutrino_Abbasi,bursts_lack_neutrino_Taboada,bursts_lack_neutrino_IceCube}.
%
%
%


\section{Conclusions}

\label{boson-concl}

The aim of this paper was to examine homogeneous self-consistent
ground state solutions
in the CGSW model  \cite{Dziekuje_Jacek_nova_2}.
It is an effective one as is the GSW model which is its quantum counterpart.
It is assumed that if the ground state of the configuration of the
self-fields induced by extended (non-bosonic) charge fluctuations appears \cite{JacekManka},
then this forces us to describe the physical system inside its droplet in the manner of classical field theory. \\
%
%
Let us summarize the results presented in this paper.
The discussed model is homogeneous on the level of one droplet (thus the thin wall approach is used).
The homogeneous configurations of the gauge ground self-fields
$W^{\pm}_{1,2}$, $Z_{0}$ and $A_{0}$
%
%
and the scalar field fluctuation $\varphi_{f}$
in the presence of a spatially extended homogeneous basic fermionic
fluctuation(s) that carries the nonzero charges
were examined.
The ground fields penetrate the whole spatially extended
fermionic fluctuation(s) and in their presence the
electroweak force generates ``the electroweak screening fluctuation of
charges'' according to Eqs.(\ref{Pdz_32})-(\ref{Pdz_35}).
\\
In general, we notice two physically different configurations of the
fields. When a matter source has the charge density fluctuation $\varrho_{f Q \; SR} \neq 0$, then classes {\it \cite{diffrent-classes} }
%
%
%
%
%
of the ground fields configurations EWbgfms (with $\vartheta \neq 0 $
and $\delta \neq 0$) that are induced by this source exist (see Section~\ref{electric}).
%
%
%
%
The mass (\ref{M od rQ}) of a droplet of this configuration of field was determined for the value of the matter electric charge fluctuation equal to $q_{f}$, (\ref{qf dla MQ}).
The EWbgfms configurations lie on the $M_{q_{f}}(\varrho_{f Q})$ curves (see Figure~4a) or equivalently on the ${\cal E}_{st}(\varrho_{f Q \; SR})$ curves only (see Figure~3b).
For the particular value of $p$,
%
%
the functions $M_{q_{f}}(\varrho_{f Q \; SR})$, (\ref{M od rQ}), and their minima $M^{min}_{q_{f}}$ depend on $\lambda$ (see Figure~4a). \\
Inside the droplet,
both the appearance of the mass of the (non self consistently treated) wavy self-field $\tilde{A}_{\mu}$ and the modification of the masses of the wavy self-fields
$\tilde{W}^{+}_{\mu}-\tilde{W}^{-}_{\mu}$, $\tilde{Z}_{\mu}$ and also the scalar fluctuation field $\varphi_{f}$ are caused due to the existence of the self-consistent fields (see Section~\ref{The screening condition}) and
%
%
due to the screening effect of the fluctuation of charges
formulated by Eqs.(\ref{Pdz_32})-(\ref{Pdz_35}).
Then, the obtained masses are used in order to estimate the thin wall approximation range.
A more complete description of the EWbgfms configurations, e.g. the dependance of the observed charge density fluctuation $\varrho_{f Q}$ on $\varrho_{f Q \; SR} \neq 0$ and the modification of the mixing angle $\Theta$, (\ref{Pdz_52}), with a change of $\varrho_{f Q}$ and the stability of the EWbgfms configurations is given in Section~\ref{electric}.
\\
%
When the weak charge density fluctuation $\varrho_{f Z \; SR} \neq 0$ (and $\varrho_{f Q \; SR} = 0$)
then the electrically uncharged, weakly charged Wbgfms configurations with $\vartheta = 0$ and
$\delta \neq 0$ and the ground self-field $Z_{0} = \zeta \neq 0$ can  exist  (see Section~\ref{neutral}).
The region of the stable (for the sake of the $W^{\pm}$ sector)  Wbgfms configurations lies on and below the $\varrho_{f Z \, SR max}(\lambda_{max})$ boundary curve (see Figure~7a). For the particular value of ${i^{3}_{f}}$, (\ref{ri3 radius of Mi3}), the function $\varrho_{f Z \, SR max}(\lambda_{max})$ gives the function $M^{max}_{i^{3}_{f}}(\lambda_{max})$, which divides the plane
$ \lambda  \times M_{i^{3}_{f}}$ of all Wbgfms configurations into the stability and instability regions (see Figure~7b). A more complete description of the Wbgfms configurations can be found in Section~\ref{neutral}.

Previously, in \cite{Dziekuje_Jacek_nova_2} it was found that for $\lambda =1$ and for $p= 2$  a shallow minimum of the mass of the EWbgfms configuration droplet equal to $M^{min}_{q_{f}} \approx \pm q_{f} \times 66.7464 \; GeV$ appears. At that time the expectation was that the appearance of such bgfms configurations might be theoretically possible in the very dense microscopic objects that are created in heavy ion collisions \cite{ion-collisions}.
%
%
%
In the present paper in Section~\ref{intersection}, the complete characteristics of two such bgfms configurations $s1_{q_{f}=2}$ and $s2_{q_{f}=2}$ were given. We only remind the reader that for the zero spin $s1_{q_{f}=2}$ state
(realized for $\lambda \approx 0.0652 $)
%
%
the mass of the EWbgfms droplet equal to $M^{min}_{q_{f} = 2} \approx 126.67$ ${\rm GeV}$ was obtained.
The physical realization of the other
EWbgfms $s2_{q_{f} = 2}$ state (at least as far as its mass is taken into account) is doubtful, as the fields configuration inside the droplet of its electrically neutral Wbgfms successor $s2_{i^{3}_{f} = \mp 1/2}$ is induced by one fermionic fluctuation only thus having a  half spin value, which is
less consistent with the observations reported in the LHC experiment \cite{spin_in_LHC_Miller_1,spin_in_LHC_Miller_2}, \cite{spin_in_LHC_Gao,spin_in_LHC_Englert,spin_in_LHC_Djouadi,spin_in_LHC_Ellis}. As it was previously mentioned only, the algebraic mean of the mass of the central
solutions $s1$ and $s2$,  i.e.,  126.67 GeV and 123.7 GeV, respectively, is equal to 125.185 GeV.
\\
%
%
Thus, the remaining, zero spin EWbgfms state $s1_{q_{f}=2}$ is the configuration in the minimum of the $M_{q_{f}}(r_{q_{f}})$ curve for $p=2$, $q_{f}=2$ and with $\lambda \approx 0.0652$ (see Figure~4a). It lies on the $M^{min}_{q_{f}=2}(\lambda)$ curve at the point of its intersection with the  boundary curve  $M^{max}_{i^{3}_{f}=\mp 1}(\lambda=\lambda_{\max} \approx 0.0652)$
(see Figure~8a).
The intersection point
%
%
is interpreted as the one that corresponds to the transition  of the electrically charged EWbgfms  configuration $s1_{q_{f}=2}$ to the electrically uncharged zero spin
Wbgfms state  $s1_{i^{3}_{f} = \mp 1}$, which has the mass $M^{max}_{i^{3}_{f}=\mp 1} \approx 126.67$ ${\rm GeV}$, as can be seen in Figure~8b.
%
%
In Section~\ref{intersection} it was argued that the configuration $s1_{i^{3}_{f} = \mp 1}$
corresponds to the LHC $\sim 126.5$ GeV zero spin state.
%
%
This physically interesting solution, which is discussed in the present paper, has  not been found before  (see Figure~8a). \\
%
%
In this paper it was also noted that for both the EWbgfms and Wbgfms  configurations the non-zero charge fluctuations (fundamentally $\varrho_{f Y}$) imply a non-zero value of the self-consistent field $\delta \neq 0$ of the scalar fluctuation  $\varphi_{f}$ (compare Notes in Section~\ref{electric} below Eq.(\ref{Pdz_56}) and in Section~\ref{neutral} below Eq.(\ref{Pdz_70})).
Thus, in the more fundamental theory, the self-consistent field $\delta$ could be a secondary quantity. Because for both EWbgfms and Wbgfms  configurations (for which $\varrho_{f Y} \neq 0$), we find that the limit  $\varrho_{f Y} \rightarrow 0$ implies $\delta \rightarrow v$ thus a derivative meaning for the parameter $v$ of the scalar fluctuation potential may also be suggested. \\
Finally,
%
%
if Wbgfms state $s1_{i^{3}_{f} = \mp 1}$ is interpreted as the LHC $\sim 126.5$ GeV one, then this means that
the value of $\lambda  = \lambda_{max} \approx 0.0652$, which is the constant parameter of the CGSW model, is a little bit bigger than the limiting stability value $\lambda^{limit} = g^{2}/(16 \; cos^{4}\Theta_{W}) \approx 0.0451$ (see Section~\ref{neutral} and Figures~7a-b). A bgfms state exists for
%
%
$\lambda \approx 0.0652$ only  (although other specific values of $\lambda $ are possible in an extension of the model, see Eq.(\ref{pairs of states})).
%
%
Therefore, a Wbgfms configuration of fields
with  $\varrho_{f Z \; SR}$ bigger than $\varrho_{f Z \; SR max} \approx 9.249 \times 10^6$ ${\rm GeV}^3$ (which is the density for $s1_{i^{3}_{f} = \mp 1}$ state, calculated in accordance with Eq.(\ref{rhomax})) lies above the $s1_{i^{3}_{f} = \mp 1}$ state in the instability region (see Figure~7), and is unstable in the $W^{\pm}$ sector.
Therefore, as was suggested in \cite{Dziekuje_Jacek_nova_2}, it radiates
%
%
to the states with $\varrho_{f Z \, SR} \leq \varrho_{f Z \, SR max}$ or decays into stable particles, i.e. photons, leptons, hadrons  and
%
%
neutrinos, as was described in Section~\ref{process}.

The non-linear self-consistent classical field theory is inherently connected with the existence of the self-field \cite{Barut-1,Barut-2,Barut-3,Barut-4}, \cite{B-Nonlinear-1,B-Nonlinear-2}, \cite{bib_B-K-1} coupled to the basic field (fluctuation).
For example, in the perturbative QED the classical self-field of the electron fluctuation is completely absent and it comes back in via a separate quantized radiation field ``photon by photon''. Meanwhile, in the self-consistent classical field concept, the whole self-field is put in from the beginning. It is free of the idea of the quantum field theory vacuum (state) \cite{Nedelko:2014sla} and the virtual pair creation.
\\
The self-field concept was  previously used with great success in the Abelian case
%
%
e.g. in order to compute nonrelativistic Lamb shifts and spontaneous
emission
\cite{spontaneous-1,spontaneous-2},
the Lamb shift (obtained iteratively) \cite{relativ_Lamb},  spontaneous emission in cavities \cite{bib_B-D} and
long-range Casimir-Polder van der Waals forces \cite{Casimir}.
%
%
%
%
%
%
%
%
%
These  analyses follow the work of Jaynes and Milonni \cite{Jaynes-1,Jaynes-2,Jaynes-3}, \cite{Milonni} and the even earlier 1951 paper of Callen and Welton \cite{Callen-Welton} on the fluctuation dissipation theorem, which showed that there is an intimate connection between vacuum fluctuations and the process of radiation reaction. The existence of one implies the existence of the other. \\
%
%
The linear Dirac equation alone with
e.g. the electron wave function in the presence of the (external to it) Coulomb field
%
%
leads to wave mechanical solutions for the ground and excited states of the electron in an atom (see Introduction).
The mathematics of the non-linear
%
%
Dirac equation for the basic field fluctuation, which follows from the coupled Maxwell and linear Dirac equations for this
%
%
fluctuation and its electromagnetic self-field is quite different.
In general,
the mathematics of the self-consistent field theory
is interested in a proper set of partial differential equations,
%
%
which are then solved self consistently in such a way that all degrees of freedom are removed.
What remains is one particular state of the
system. \\
{\bf Remark}: For example,
the self-consistent solution of the
%
%
couple: the Dirac equation and classical Maxwell equations will give
a real photon that is a ``lump of electromagnetic substance'' (without Fourier decomposition \cite{Dziekuje_za_channel,Dziekuje_za_skrypt} as is suggested from recent experiments \cite{Roychoudhuri})
as the reflection of the coupling to the Dirac equation. If we pull back from this particular solution forgetting about the primary Dirac equation then what remains are not the classical Maxwell equations for the classical electromagnetic field but
%
%
equations that act on
the space of possible photonic states. QED with the field operator and the Fock space have to be the non-self-consistent reflection of this construction (if only the Fourier decomposed frequencies of the light  pulse represent actual optical frequencies, which has
recently been questioned  by  light beam experiments \cite{Roychoudhuri}).
%
%
(Compare the self-consistent pair of equations (\ref{Pdz_69})-(\ref{Pdz_70}) with the non-self-consistent Eq.(\ref{delta from i3})).
\\
The merits of the thought that is
behind this procedure is the self consistency of the solution. The further we are from this precise self-consistent solution, the more numerous a set of differential equations remains to be solved but the set of equations that are already solved determines  the types of the equations which remain and the properties of the fields that are ruled by them.  \\
The self-field is small for atomic phenomena and therefore the description of the basic field fluctuation via the linear Dirac equation may work approximately, which follows from the fact that the non-linear terms are small and can be treated as perturbations.
%
%
Nevertheless, the QED prevailed, mainly because of the successes in the scattering phenomena.  \\
Yet, the self-field is not always small and there is another region where
%
%
the non-linear terms dominate \cite{B-Nonlinear-1,B-Nonlinear-2}.
The present paper reflects  such a situation, since for the bgfms configuration of fields,  the energy of the host
%
%
fermionic fluctuation is assumed to be minute in comparison to the obtained mass of the bgfms  droplet.
Thus, the main theoretical subject of this paper was the self-consistent description of the configuration of electroweakly interacting self-fields that are induced by a charge density fluctuation(s) with the internal extended wave structure inside one droplet. Thus, the CGSW model is the type of ``a source theory'' that  considers all self-fields and scalar field fluctuations as ``derived'' from the source of the fluctuations of charges. The quotation marks mean that the self-consistent fields are not absent - they are only self consistently derived from the basic fluctuations fields to which they are coupled via the screening condition of the fluctuation of charges (\ref{Pdz_32})-(\ref{Pdz_35}).
%
%

In the presented CGSW model of the bgfms configuration of fields induced by the basic matter field fluctuation(s), the droplet is like the whole particle. This is connected with the fact that (besides the fact that the energy of the fermionic fluctuation is ignored) any fermionic fluctuation which ``stretches'' the droplet is like
a whole fermion. Thus, our droplet of the bgfms configuration is like ``a parton''. This is definitely not the most general case. \\
The indispensable need for the development of a more general approach is seen from
%
%
the self-consistent model of the configuration of fields induced by the electronic charge fluctuation used in the Lamb shift explanation, where the energy of the electronic
%
%
fluctuation is ignored (not to mention the ground and excited states of the electron, which are obtained in the  anticipation by the formalism of the wave mechanics for the total electron wave function that is  treated non-self consistently).
Therefore, let us assume that there is an object in which the fluctuation of the fermionic charge does not exist by itself but needs a globally extended fermionic charge of which it is the disturbance only. With such approach, one is obliged to define and find the mass of the configuration of fields induced by the globally extended charge together with its fluctuation(s) (extended globally or locally). In doing this, one should focus on neither the wave mechanics (or quantum mechanics) nor on the self-consistent field theory of fluctuations (or quantum field theory) but on the theory of the complete inner structure of one particle. Otherwise, the model gets into
the composition of ``a particle'' from ``partons'', which is a kind of  ``planetarianism'' and seemingly because of this e.g. quantum chromodynamics (QCD) is the theory without final fundamental success \cite{nucleac_spin_crisis}, as  was expressed in \cite{Heyde}: ``... all spin parts $\left[ {\rm of \; the \;
nucleon} \right] $ have to add to $\frac{1}{2}$ which is
incredible in the light of the present day experiments. This may
indicate that some underlying symmetries, unknown at present, are
playing a role in forming the various contributing parts such that
the final sum rule gives the fermion $\frac{1}{2}$ value''.\\
%
%
Both to recapitulate and going a little bit further, in order to describe the state of one particle (or even one droplet with a fluctuation) in a fully self-consistent way, the interaction of the self-fields with the globally extended charge and fluctuations inside this particle (possibly ruled by equations unknown at present) has to be considered simultaneously.
Consequently, further analysis should describe a more realistic shape of the
%
%
charge density
%
%
of the extended matter source. Supposing that proper equations are known, this shape should follow e.g. from the coupled Klein-Gordon-Maxwell
(Yang-Mills) or Dirac-Maxwell (Yang-Mills) equations \cite{Ilona}
%
%
and from the Einstein's equations (or equations of an effective gravity theory of the Logunov type \cite{Denisov-Logunov,Lammerzahl}) as is required for the self-consistent models. Thus, to make the theory of one particle fully self-consistent even a
model of gravitation should be included \cite{Dziekuje_za_neutron}.
%
%
Hence, a matter particle (similar to one droplet induced by matter fluctuations) seems to be, from the
mathematical point of view, a self-consistent solution of all of the field equations involved in the description of the constituent fields inside this particle. Its interaction as a whole with the outer world is ruled by other models. \\
The presented electroweak CGWS model, although elaborated on for configurations of fields inside one particle that are induced by the basic matter fluctuations only, is the next step towards the self-field formalism
\cite{Dziekuje_za_skrypt}, \cite{Frieden-1,Frieden-2,Frieden-3,Frieden-4,Frieden-5}, \cite{Dziekuje_za_models_building,Dziekuje_informacja_2,Dziekuje_informacja_1,Dziekuje_za_channel}
of the classical theory of one elementary particle. This particle is a materially extended entity with its own self-fields (e.g.
electroweak, gravitational, etc.) coupled self consistently to the basic fields inside it.
In \cite{Dziekuje_za_neutron} and in the present paper, it is suggested that the realization of such an analysis in the derivation of the characteristics of one particle is at hand.

\vspace{-2mm}

\begin{acknowledgments}
This work has been supported by L.J.CH..\\
This work has been also supported by the Department of Field
Theory and Particle Physics, Institute of Physics, University of
Silesia and by the Modelling Research Institute, 40-059 Katowice,
Drzyma{\l}y 7/5, Poland.
\end{acknowledgments}


\vspace{3mm}

\begin{center}

\begin{widetext}

\vspace{2mm}

\section{Appendix~1: Quantum numbers in the CGSW model} \nopagebreak[4]

\label{boson-tabl}

\ \ \ \ \ \ \ Table: Some quantum numbers in the CGSW $SU_{L}(2) \times U_{Y}(1)$ model.
\nopagebreak[4]

\begin{tabular}{ccccc}   \hline

           & Weak           & Weak         & Electric          &    \\
           & Isotopic       & Hypercharge  & Charge $Q$        &$p = 2Q/Y$ \\
           & Charge $I^{3}$ & $Y$          & $ Q = I^{3} +Y/2$ &     \\
\hline

$leptonic \; fluctuations$         &         &              &                   &     \\

$\nu_{f \,L}$         &    1/2  &  - 1         &    0              &   0 \\
$ \ell_{f \, L}  $        &  - 1/2  &  - 1         &  - 1              &   2 \\
                  &         &              &                   &     \\
$\ell_{f \, R}$           &     0   &  - 2         &  - 1              &   1 \\
$\ell = e, \mu, \tau$ &         &              &                   &     \\
\hline

$gauge \; self \; fields$ &         &              &                   &     \\

$W^{+}$           &     1   &    0         &    1              &     \\
$W^{3}$           &     0   &    0         &    0              &     \\
$W^{-}$           &  -  1   &    0         &  - 1              &     \\
                  &         &              &                   &     \\
$ B $             &     0   &    0         &    0              &     \\
\hline

$scalar \; fluctuations$  &         &              &                   &     \\
$doublet \; \Phi_{f}$  &         &              &                   &     \\

$ \Phi_{f}^{+} $         &    1/2  &    1         &    1              &   2 \\
$ \Phi_{f}^{0}$          &  - 1/2  &    1         &    0              &   0 \\
\hline
                  &         &              &                   &          \\
$some$            &  - 1/2  &    1         &    0              & 0      \\
$source$  &  - 1    &    4         &    1              &   1/2   \\
$matter$          &    0    &    2         &    1              &   1    \\
$fluctuation$    &   1/2   &    1         &    1              & 2
\\
$configurations$                  &   3/2   &    1         &    2              &   4  \\
                  &         &              &                   &        \\
\hline

\end{tabular}


\vspace{10mm}

\end{widetext}

\vspace{5mm}

\end{center}


\newpage

\section{Appendix~2: The CGSW model field equations with continuous matter current density fluctuations}

\label{appendix}

From (\ref{lagrangian}) the field equations for the Yang-Mills self fields follow
$(\Box = \partial_{\nu} \partial^{\nu})$, for $B^{\mu}$
\begin{eqnarray}
\label{Pdz_14}
&-& \Box B^{\mu} + \partial^{\mu}\partial_{\nu}B^{\nu} = \nonumber \\
&=& - \frac{1}{4}gg'\varphi_{f}^{2}W^{3 \mu} +
\frac{1}{4}g'^{2}\varphi_{f}^{2}B^{\mu} - \frac{g'}{2} j_{f \,
Y}^{\, \mu} \; ,
\end{eqnarray}
for $W^{a \mu} (a = 1,\; 2)$
\begin{eqnarray}
\label{Pdz_15}
&-& \! \Box W^{a \mu}  +  g\varepsilon_{a b c}W^{b \nu}\partial_{\nu}W^{c \mu} =
 \\
\!\!\!\!\!\!\! & = & \! g^{2}(\frac{1}{4}\varphi_{f}^{2}W^{a \mu} - W^{b}_{\nu}W^{b
\nu}W^{a \mu} + W^{a \nu}W^{b}_{\nu}W^{b \mu}) - g j_{f}^{\, a
\mu} \, , \nonumber
\end{eqnarray}
and for $W^{3 \mu}$
\begin{eqnarray}
\label{Pdz_16}
& - & \! \Box W^{3 \mu}  +  g\varepsilon_{3 b c}W^{b
\nu}\partial_{\nu}W^{c \mu}
=  \frac{1}{4}g^{2}\varphi_{f}^{2}W^{3
\mu}
-  \frac{1}{4}gg'\varphi_{f}^{2}B^{\mu}
\nonumber  \\
& - & \!
g^{2}W^{b}_{\nu}W^{b \nu}W^{3 \mu} + g^{2}W^{3 \nu}W^{b}_{\nu}W^{b
\mu} - g j_{f}^{\, 3 \mu} \, .
\end{eqnarray}
Here $j_{f \, Y}^{\mu}$ and $j_{f}^{\, a \mu}$
are the continuous matter current density fluctuations extended
in space, which are given by equations
\begin{eqnarray}
\label{Pdz_17} j_{f \, Y}^{\, \mu} =
\overline{L_{f}}\gamma^{\mu}YL_{f} +
\overline{R_{f}}\gamma^{\mu}YR_{f} \; ,
\end{eqnarray}
\begin{eqnarray}
\label{Pdz_18} j_{f}^{\, a \mu} =
\overline{L_{f}}\gamma^{\mu}\frac{\sigma^{a}}{2}L_{f} \; , \; \;\;\; {\rm
where} \;\;\; a = 1,2,3 \; .
\end{eqnarray}
Similarly, the fluctuation $\varphi_{f}$
of the scalar field satisfies
\begin {eqnarray}
\label{Pdz_19}
-  \Box \varphi_{f}  &=&
(-\frac{1}{4}g^{2}W^{a}_{\nu}W^{a \nu} -
\frac{1}{4}g'^{2}B_{\nu}B^{\nu} + \frac{1}{2}
gg'W^{3}_{\nu}B^{\nu})\varphi_{f} \nonumber \\
&-& \!   \lambda v^{2}\varphi_{f} + \lambda\varphi_{f}^{3} +
\frac{m_{\ell_{f}}}{v}(\overline{\ell}_{f  L} \, \ell_{f  R} + \;h.c.) \;
.
\end{eqnarray}
To simplify the calculations, we neglect the mass $m_{\ell_{f}}$ of
the fermionic fluctuation $\ell_{f}$. It could be smaller than the mass of e.g. electron. But, if  $\ell_{f}$ would coincide with the lepton $\ell$, e.g. electron, then it enters with a relative strength equal to $\frac{m_{e_{f}}}{v} \sim 2.1 \times 10^{-6} $.

\end{document}